# The Imaging X-Ray Polarimetry Explorer (IXPE): Pre-Launch

Martin C. Weisskopf [1a], Paolo Soffitta [8], Luca Baldini [4,3], Brian D. Ramsey [1], Stephen L. O'Dell [1], Roger W. Romani [18], Giorgio Matt [11], William D. Deininger [5], Wayne H. Baumgartner [1], Ronaldo Bellazzini [3], Enrico Costa [8], Jeffery J. Kolodziejczak [1], Luca Latronico [9], Herman L. Marshall [10], Fabio Muleri [8], Stephen D. Bongiorno [1], Allyn Tennant [1], Niccolo Bucciantini [29], Michal Dovciak [30], Frederic Marin [28], Alan Marscher [31], Juri Poutanen [32], Pat Slane [33], Roberto Turolla [34], William Kalinowski [5], Alessandro Di Marco [8], Sergio Fabiani [8], Massimo Minuti [3], Fabio La Monaca [8], Michele Pinchera [3], John Rankin [8,16], Carmelo Sgro' [3], Alessio Trois [13], Fei Xie [8], Cheryl Alexander [19], D. Zachery Allen [5], Fabrizio Amici [8], Jason Andersen [5], Angelo Antonelli [14], Spencer Antoniak [5], Primo Attina' [2], Mattia Barbanera [3], Matteo Bachetti [13], Randy M. Baggett [1], Jeff Bladt [5], Alessandro Brez [3], Raffaella Bonino [9,20], Christopher Boree [5], Fabio Borotto [9], Shawn Breeding [1], Daniele Brienza [8], H. Kyle Bygott [5], Ciro Caporale [9], Claudia Cardelli [3], Rita Carpentiero [6], Simone Castellano [3], Marco Castronuovo [6], Luca Cavalli [7], Elisabetta Cavazzuti [6], Marco Ceccanti [3], Mauro Centrone [14], Saverio Citraro [3], Fabio D' Amico [6], Elisa D'Alba [7], Laura Di Gesu [6], Ettore Del Monte [8], Kurtis L. Dietz [1], Niccolo' Di Lalla [18], Giuseppe Di Persio [8], David Dolan [5], Immacolata Donnarumma [6], Yuri Evangelista [8], Kevin Ferrant [5], Riccardo Ferrazzoli [8], MacKenzie Ferrie [5], Joseph Footdale [5], Brent Forsyth [12], Michelle Foster [1], Benjamin Garelick [5], Shuichi Gunji [21], Eli Gurnee [5], Michael Head [5], Grant Hibbard [5], Samantha Johnson [1], Erik Kelly [5], Kiranmayee Kilaru [22], Carlo Lefevre [8], Shelley Le Roy [23], Pasqualino Loffredo [8], Paolo Lorenzi [7], Leonardo Lucchesi [3], Tyler Maddox [5], Guido Magazzu [3], Simone Maldera [9], Alberto Manfreda [3], Elio Mangraviti [7], Marco Marengo [9], Alessandra Marrocchesi [3], Francesco Massaro [9,20], David Mauger [5], Jeffrey McCracken [19], Michael McEachen [24], Rondal Mize [1], Paolo Mereu [9], Scott Mitchell [5], Ikuyuki Mitsuishi [25], Alfredo Morbidini [8], Federico Mosti [9], Hikmat Nasimi [3], Barbara Negri [6], Michela Negro [9,20], Toan Nguyen [5], Isaac Nitschke [5], Alessio Nuti [3], Mitch Onizuka [5], Chiara Oppedisano [9], Leonardo Orsini [3], Darren Osborne [12], Richard Pacheco [5], Alessandro Paggi [9,20], Will Painter [5], Steven D. Pavelitz [1], Christina Pentz [5], Raffaele Piazzolla [6], Matteo Perri [14], Melissa Pesce-Rollins [3], Colin Peterson [5], Maura Pilia [13], Alessandro Profeti [3], Simonetta Puccetti [15], Jaganathan Ranganathan [1], Ajay Ratheesh [8,17], Lee Reedy [12], Noah Root [5], Alda Rubini [8], Stephanie Ruswick [12], Javier Sanchez [23], Paolo Sarra [7], Francesco Santoli [8], Emanuele Scalise [8], Andrea Sciortino [7], Christopher Schroeder [5], Tim Seek [5], Kalie Sosdian [5], Gloria Spandre [3], Chet O. Speegle [1], Toru Tamagawa [26], Marcello Tardiola [9], Antonino Tobia [8], Nicholas E. Thomas [1], Robert Valerie [5], Marco Vimercati [7], Amy L. Walden [1], Bruce Weddendorf [27], Jeffrey Wedmore [5], David Welch [12], Davide Zanetti[3], Francesco Zanetti [7]

[1] NASA Marshall Space Flight Ctr., Huntsville, AL 35812, USA
[2] INAF Osservatorio Astrofisico di Torino, Strada Osservatorio 20, 10025 Pino Torinese (TO), Italy
[3] Istituto Nazionale di Fisica Nucleare - Pisa, Largo Bruno Pontecorvo 3, 56127 Pisa (PI), Italy
[4] Universita' di Pisa, Lungarno Antonio Pacinotti 43, 56126 Pisa (PI), Italy
[5] Ball Aerospace, 1600 Commerce Street, Boulder, CO 80301 USA
[6] Agenzia Spaziale Italiana, Via del Politecnico, 00133 Roma (RM), Italy
[7] OHB Italia SpA, Via Gallarate 150, 20151 Milano (MI), Italy

[8] INAF Istituto di Astrofisica e Planetologia Spaziali, Via del Fosso del Cavaliere 100, 00133 Roma (RM), Italy
[9] Istituto Nazionale di Fisica Nucleare - Torino, Via Pietro Giuria 1, 10125 Torino (TO), Italy
[10] Massachusetts Institute of Technology, 77 Massachusetts Avenue, Cambridge, MA 02139, USA
[11] Universita' degli Studi Roma Tre, Via della Vasca Navale 84, 00146 Roma (RM), Italy
[12] Lab. for Atmospheric and Space Physics, 1234 Innovation Drive, Boulder, CO 80303, USA
[13] INAF Osservatorio Astronomico di Cagliari, Via della Scienza 5, 09047 Selargius (CA), Italy
[14] INAF Osservatorio Astronomico di Roma, Via Frascati 33, 00078 Monte Porzio Catone (RM), Italy
[15] ASI Space Science Data Ctr., Via del Politecnico, Edificio D, 00133 Roma (RM), Italy
[16] Universita' di Roma "La Sapienza", Piazzale Aldo Moro 5, 00185 Roma (RM) Italy
[17] Universita' degli Studi di Roma "Tor Vergata", Via Lucullo, 11, 00187 Roma (RM), Italy
[18] Stanford University, 382 Via Pueblo Mall, Stanford, CA 94305-4060, USA
[19] Linc Research, Marshall Space Flight Ctr., Huntsville, AL 35812, USA
[20] Universita' degli Studi di Torino, Via Giuseppe Verdi 8, 10124 Torino (TO), Italy
[21] Yamagata University, 1-4-12 Kojirakawa-machi, Yamagata-shi, 990-8560, Japan
[22] USRA, Marshall Space Flight Ctr., Huntsville, AL 35812, USA
[23] ESSCA, Marshall Space Flight Ctr., Huntsville, AL 35812, USA
[24] Northrop Grumman Innovation Systems, 600 Pine Avenue, Goleta, CA 93117, USA
[25] Nagoya University, Furo-cho, Chikusa-ku, Nagoya, 464-8602, Japan
[26] RIKEN Nishina Ctr., 2-1 Hirosawa, Wako, Saitama 351-0198, Japan
[27] Weddendorf Design, 14060 Valley Vista Drive, Huntsville, AL 35803, USA
[28] Université de Strasbourg, CNRS, Observatoire Astronomique de Strasbourg, UMR 7550, 67000 Strasbourg, France
[29] INAF Osservatorio Astrofisico di Arcetri, Largo Enrico Fermi 5, 50125 Firenze, Italy
[30] Astronomical Institute, Bocni II 1401, CZ-14100 Prague, Czech Republic
[31] Department of Astronomy, Boston University, 725 Commonwealth Avenue, Boston, Massachusetts 02215, USA
[32] Department of Physics and Astronomy, FI-20014 University of Turku, Finland; Space Research Institute of the Russian Academy of Sciences, Profsoyuznaya str. 84/32, 117997 Moscow, Russia
[33] Center for Astrophysics, Harvard & Smithsonian, 60 Garden Street, Cambridge, Massachusetts 02138, USA
[34] Dipartimento di Fisica e Astronomia, Università di Padova, Via Marzolo 8, 35313 Padova, Italy

## ABSTRACT

Scheduled to launch in late 2021, the Imaging X-ray Polarimetry Explorer (IXPE) is a NASA Small Explorer Mission in collaboration with the Italian Space Agency (ASI). The mission will open a new window of investigation – imaging X-ray polarimetry. The observatory features 3 identical telescopes each consisting of a mirror module assembly with a polarization-sensitive imaging X-ray detector at the focus. A coilable boom, deployed on orbit, provides the necessary 4-m focal length. The observatory utilizes a 3-axis-stabilized spacecraft which provides services such as power, attitude determination and control, commanding, and telemetry to the ground. During its 2-year baseline mission, IXPE will conduct precise polarimetry for samples of multiple categories of X-ray sources, with follow-on observations of selected targets.

## 1. INTRODUCTION

The Imaging X-Ray Polarimetry Explorer[1] (IXPE) is a NASA small explorer mission selected in early 2017. IXPE will be a pathfinder mission opening a new window on the X-ray sky by enabling polarimetry measurements on essentially all classes of the brightest cosmic X-ray sources. Scheduled for launch in late 2021, with a 2-year baseline mission, IXPE will perform a study of dozens of sources in its first year, with follow-on more detailed observations of selected targets in year two.

The following overview provides a detailed technical description of the optics and the detectors, the results of the on-ground calibration utilizing both polarized and unpolarized X-ray sources and a description of the Ball-Aerospace-provided spacecraft with emphasis on those features which facilitate observations. The ground network description follows which includes the ASI-provided ground station at Malindi, the Mission Operations Center at the Laboratory for Atmospheric and Space Physics of the University of Colorado and the Science Operations Center at NASA/MSFC. NASA's HEASARC will be used for data archiving and conducting any General Observer programs in year 3 and beyond. Finally, we conclude with a brief description of some of the science that will be accomplished.

## 2. PROGRAM OVERVIEW

IXPE will be launched on a Falcon-9 rocket from Kennedy Space Center in late 2021. It will be inserted into an equatorial orbit at a nominal inclination of 0° and a nominal altitude of 600 km. The inclination minimizes the amount of time spent in the South-Atlantic Anomaly and therefore provides for a low charged-particle background. The orbit also allows frequent data downloads to the primary ground station in Malindi, Kenya (the mission will also use a backup ground station in Singapore). The chosen altitude maximizes the orbit lifetime, while still satisfying a NASA requirement of re-entry within 25 years.

The mission follows a simple observing paradigm: pointed viewing of known X-ray sources over multiple orbits (not necessarily consecutive orbits) until the observation is complete. Polarimetry, like spectroscopy, requires long integration times and, depending on the target, data collection times range from hours to many days.

Due to orientation limits on the solar panels, science targets are typically visible over an approximately 50-day window, twice a year, and can be observed continuously for a minimum time of 57 minutes each orbit depending on the target's inclination to the ecliptic plane.

Commissioning operations are conducted during the first 30 days on orbit. After separation from the Falcon 9, the Observatory autonomously detumbles, deploys the solar arrays, and performs solar acquisition. Payload commissioning includes boom deployment[2], and instrument activation. There is a Tip-Tilt-Rotate (TTR) mechanism on board which, if necessary, may be used to adjust the alignment between the optics and the detectors. Telescope (optics plus detectors) calibration activities follow, including pointing at several bright X-ray sources.

## 3. INSTITUTIONS AND ROLES

IXPE is a NASA Small Explorer mission undertaken by NASA in partnership with the Italian Space Agency (Agenzia Spaziale Italiana, ASI). The Explorer's Program Office at the Goddard Space Flight Center (GSFC) manages NASA's Explorer Program. NASA's Small Explorers are Principal Investigator (PI)-led missions: Dr. Martin C. Weisskopf of the Marshall Space Flight Center (MSFC) is the PI; Dr. Brian D. Ramsey (MSFC) the Deputy PI (DPI); Dr. Paolo Soffitta, the Italian PI; and Dr. Luca Baldini, the Italian Co-PI. IXPE project management, system engineering, and safety and mission assurance oversight are all at MSFC. In addition, MSFC was responsible for the Mirror Module Assembly (MMA) fabrication, testing, and calibration including preparation of an X-ray test facility. MSFC is also responsible for the Science Operations Center (SOC), which performs science operations and data processing. The SOC will archive IXPE data and data products at NASA's High-Energy Astrophysics Science Archive Center (HEASARC) at GSFC within 30 days of the end of an observation made during the first 3 months of operation, and subsequently within 1 week.

IXPE also has a substantial contribution from the Italian Space Agency (ASI), which has been responsible for the entire Italian contribution, managing national activities and ensuring the compliance of the Instrument (defined as the polarization-sensitive detectors and their computer) performance to requirements. This contribution included building, testing, and calibrating the Instrument including its software. Italian contributions also included support to the integration and testing of the Instrument on the satellite bus and support to the telescope calibration activities at MSFC. Italy will also support commissioning, monitoring, and maintenance of the Instrument during operations.

ASI also is providing use of its Malindi ground station for communication with the Observatory and its Fucino Space Center network, for linking to the IXPE Mission Operations Center (MOC). In addition, the ASI Space Science Data Center (SSDC) provided portions of the Calibration Database and, in collaboration with the other Italian institutions, designed and developed scientific software modules for the Instrument Pipeline at the SOC.

The Istituto di Astrofisica e Planetologia Spaziali (INAF-IAPS) runs the Italian project office which manages the Instrument activities. INAF-IAPS designed, built, and tested the on-board calibration and filter system and the two Instrument calibration stations used to accept the Gas Pixel Detectors and to calibrate the detector units (DUs). INAF-IAPS and Osservatorio Astronomico di Cagliari (INAF-OAC) also tested the Instrument computer (the Detector Service Unit, DSU) and its integration with the DUs and with the spacecraft computer. INAF-OAC leads the flight operation of the Instrument.

The Istituto Nazionale di Fisica Nucleare (INFN) designed, assembled, and delivered the DUs, including the polarization-sensitive Gas Pixel Detectors, the associated readout electronics, the housing, and the thermal system. The INFN team implemented the test equipment and associated software for the characterization and the verification of the DUs through the qualification cycle, as well as creating the full detector simulation and track-reconstruction framework in support of the calibration and the processing pipeline.

INAF and INFN remotely participated in the Observatory environmental tests at Ball Aerospace and in the X-ray calibration of the Spare Mirror and Spare Detector Unit at MSFC and the subsequent data analysis.

OHB-Italia designed, fabricated, and tested the DSU, the Filter and Calibration Wheels, the power supply, and the Electrical Ground Support Equipment. OHB also fabricated the DU back-end electronics and supported the observatory environmental test.

Università Roma Tre co-chairs the activities of the Science Working Group and of the Science Advisory Team (SAT). Specifically, it coordinates seven topical working groups for the definition of the observing program and for preparation for data analysis

The IXPE spacecraft (S/C) is provided by Ball Aerospace, which is also responsible for assembly, integration, and testing of the Observatory. In addition, Ball Aerospace manages mission operations through a subcontract with the University of Colorado’s Laboratory for Atmospheric and Space Physics (LASP).

Nagoya University (Japan) contributed thermal shields for the MMAs. RIKEN (Japan) contributed the Gas Electron Multipliers (GEMs) for the Gas Pixel Detectors.

Both Stanford University and the Massachusetts Institute of Technology have provided Co-Investigators (Co-I) as members of the IXPE Science Working Group and are involved in various aspects of the scientific mission. The Stanford Co-I co-chairs the SAT; the MIT Co-I co-chairs the Science Analysis and Simulation Working Group.

Finally, IXPE’s voluntary Science Advisory Team comprises more than 90 scientists from 12 countries.

## 4. IXPE OBSERVATORY OVERVIEW

Figure 1 shows a schematic of the IXPE observatory and its key payload elements. Pictures of the observatory are shown in Figures 2 and 3. The payload consists of three identical X-ray telescopes each comprised of a Mirror Module Assembly (MMA) with a polarization sensitive DU at its focus. Data from the detectors are handled by the DSU, located under the S/C top deck. The DSU packages the data for the S/C computer which then provides for transmission to the ground. A lightweight coilable boom deploys after launch to establish the appropriate focal length and to position each MMA above its respective detector, as the MMAs are not aligned with the DUs in the stowed position. Fixed X-ray shields (see Figure 4), in combination with collimators on each detector, limit stray radiation so that only X-ray photons that enter through an MMA can impinge on the detector entrance window. Two star trackers, one along the +Z axis as shown in Figure 1, and one pointing along the -Z axis but hidden by the S/C in the figure, provide pointing knowledge for the three-axis-stabilized spacecraft.

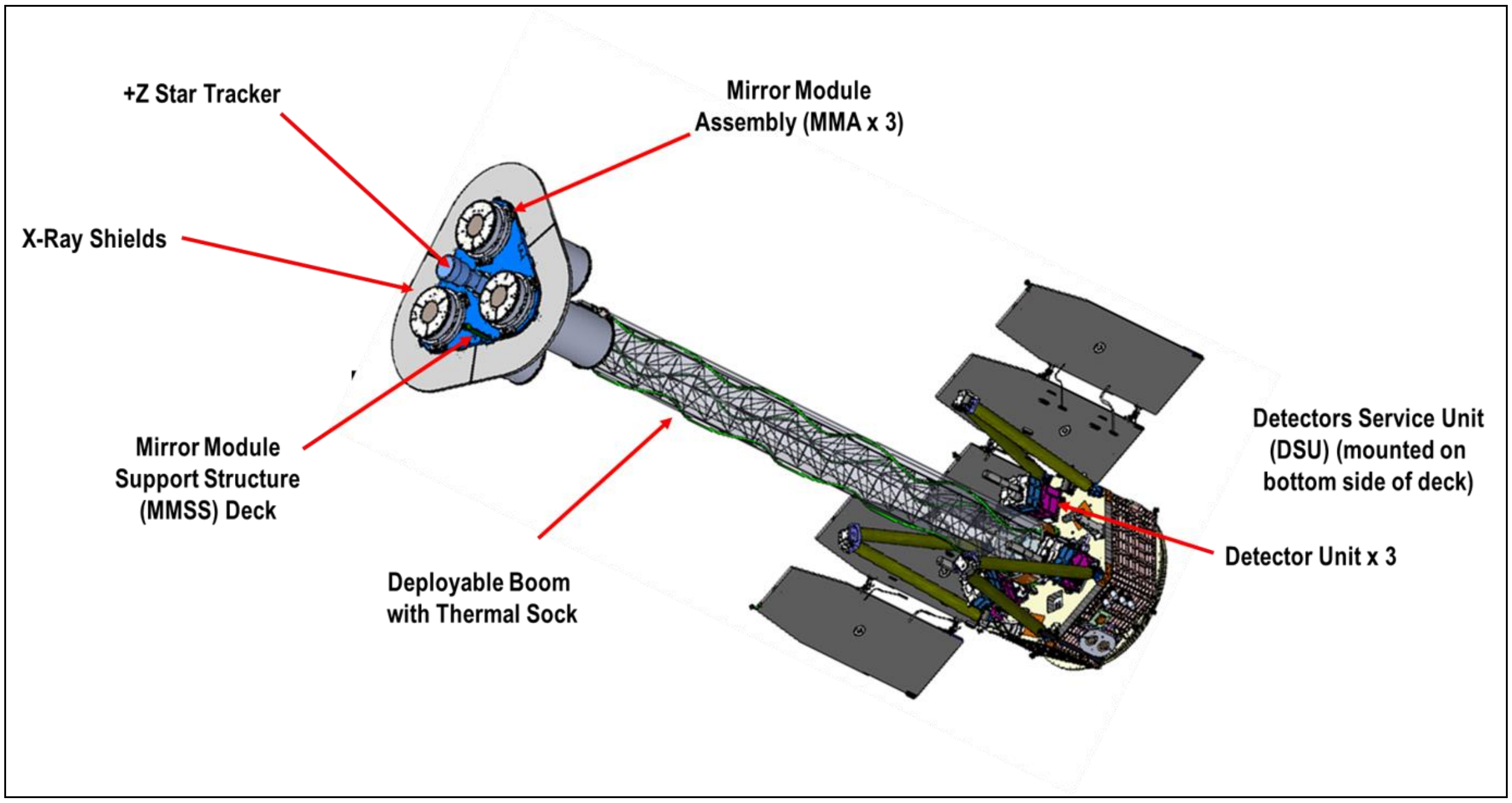

**Figure 1:** The IXPE Observatory highlighting the key scientific payload elements. A second star tracker (not visible) is on the back of pointing along the -Z axis. The DSU is a computer that provides the interface between the detector electronics and the S/C computer.

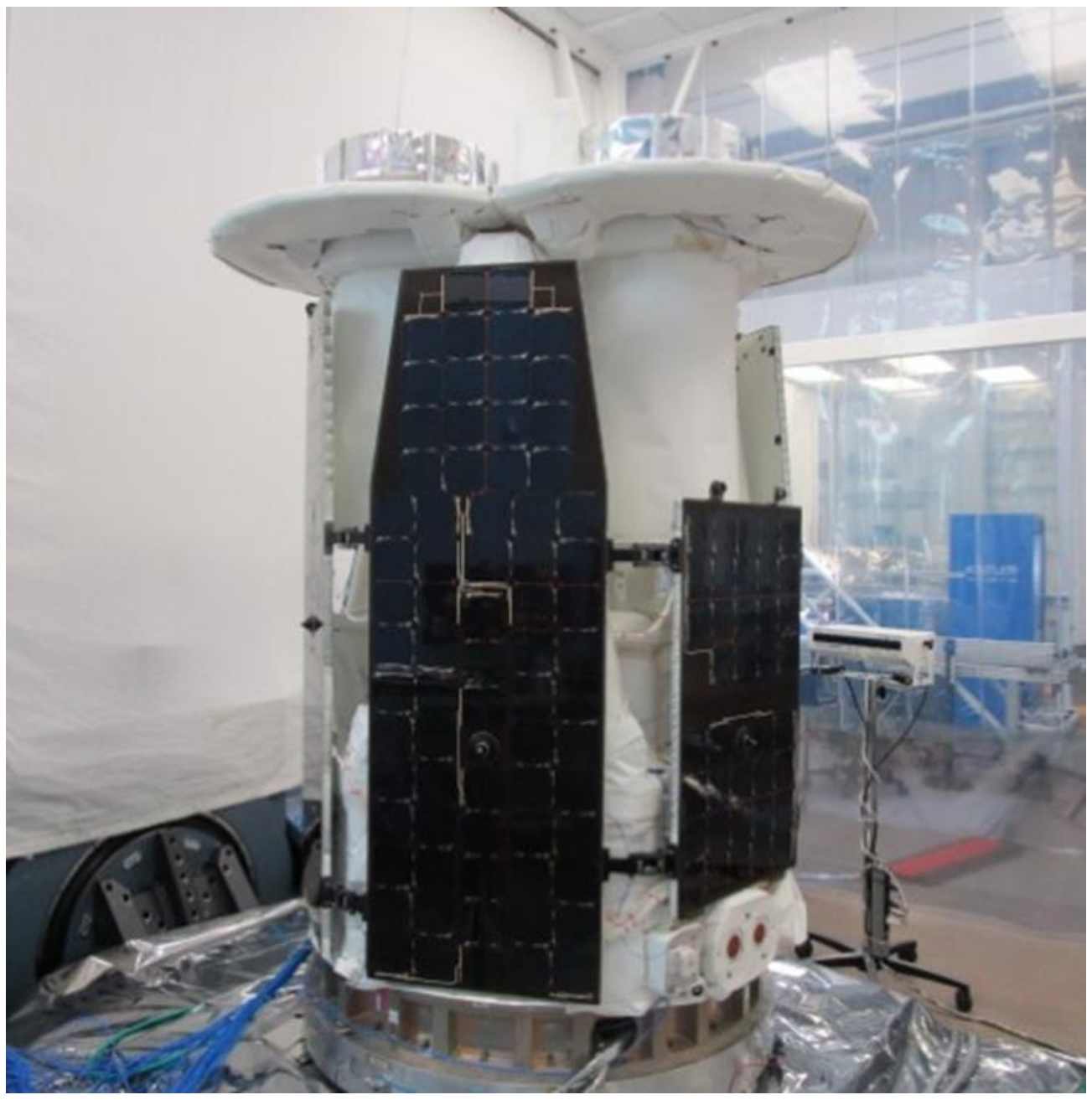

**Figure 2:** Photograph of the IXPE Observatory in the stowed position on a vibration table during Observatory environmental testing. Photo courtesy of Ball Aerospace.

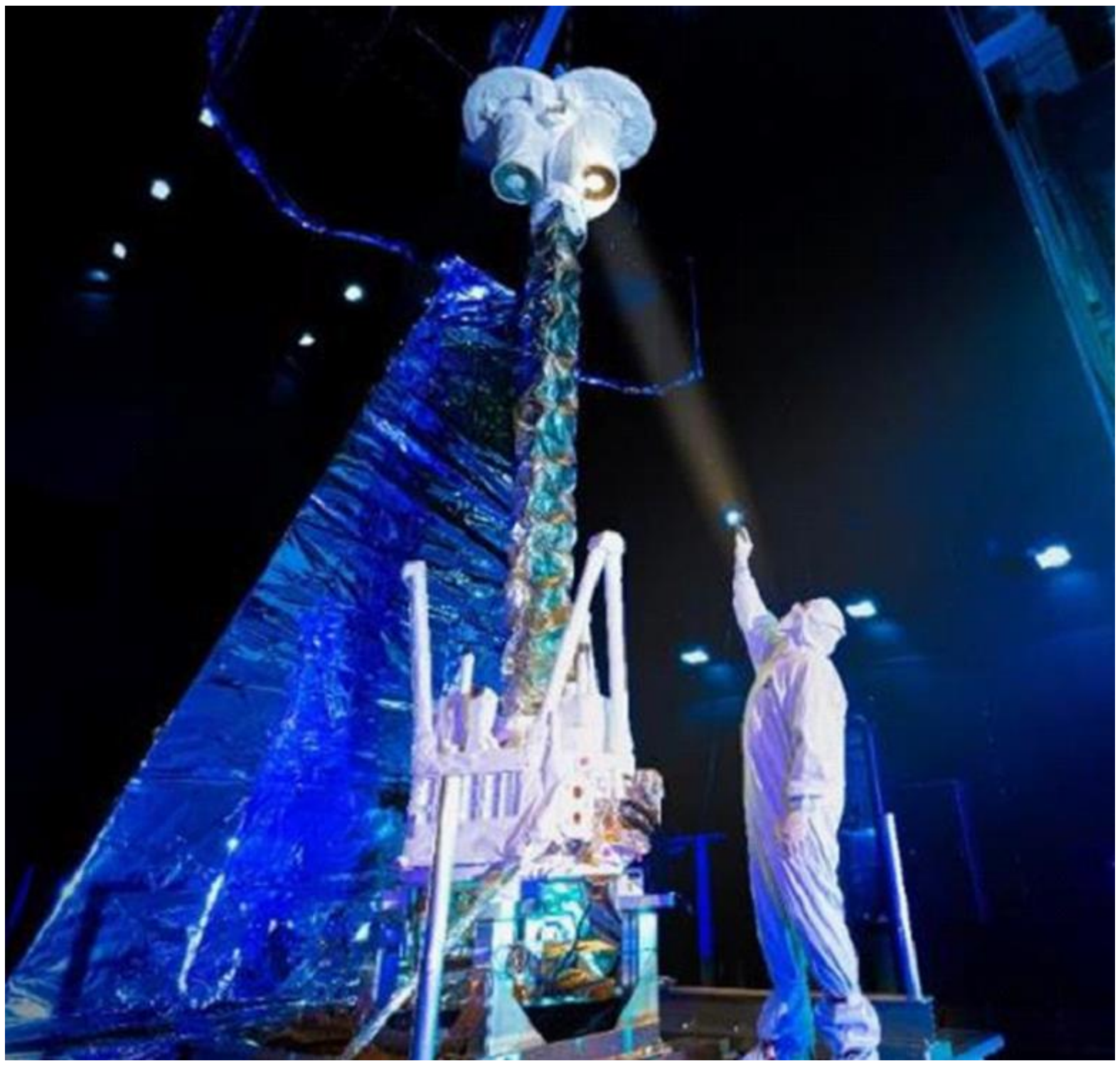

**Figure 3:** Photograph of IXPE with the boom deployed in a thermal vacuum chamber (TVAC) during Observatory environmental testing. When deployed, IXPE is 5.2 m from the bottom of the Spacecraft structure to the top of the payload and is 1.1 m in diameter. The solar panels, removed for TVAC, span 2.7 m when deployed. Photo courtesy of Ball Aerospace.

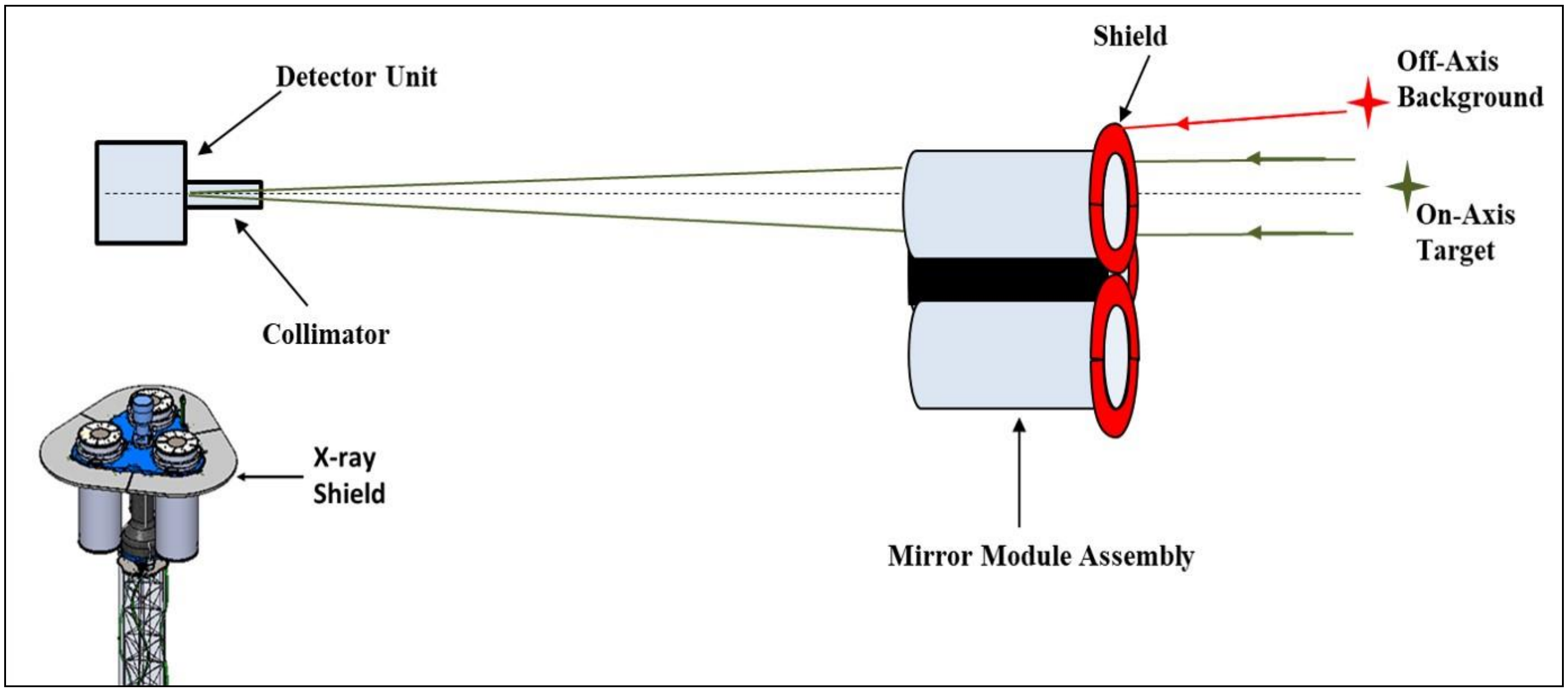

**Figure 4:** The combination of the forward X-ray shields and the collimator mounted to each detector prevents X rays, other than those passed through the mirror assemblies, from entering the detectors.

## 5. THE SPACECRAFT

The IXPE Observatory is based on the Ball Aerospace's BCP small-spacecraft architecture [3, 4]. The spacecraft supports the IXPE payload mounted on its top deck and uses a hexagonal structure with side panels to provide direct launch-load paths to the launch attach fitting and to provide surface area for spacecraft and payload components. The stowed solar array wraps around the spacecraft body enveloping the payload during launch, prior to deployment. Table 1 highlights the capabilities of the spacecraft.

The observatory attitude determination and control system (ADCS) points the science aperture at the science targets while maintaining solar array pointing at the sun. [5] For coarse attitude sensing the IXPE spacecraft utilizes an array of 12 sun sensors and a 3-axis magnetometer. For fine attitude there are two star trackers, one on the mirror module support structure, co-aligned with the axes of the MMAs (see Figure 1), and one underneath the spacecraft and pointing in the -Z direction. Pointing control is via 3 reaction wheels with torque rods used to de-saturate the wheels as necessary.

Command and data handling are controlled by an integrated avionics unit. This contains the flight software and handles the telemetry, data storage and overall payload control. Communication is via S band, with a 2-kbps command rate and a 2 Mbps downlink telemetry rate. There is 6 GBytes of on-board memory assigned for data storage between downloads.

To fit within the original launch vehicle fairing and yet provide the necessary 4-m focal length, IXPE utilizes an extending boom that deploys on orbit. The boom, provided by Northrop Grumman Space Systems, is triangular in section, with three glass-fiber longerons that extend along its full length. Battens and diagonals complete the structure and provide additional stiffness.

The boom is initially coiled in a canister and, when released, uncoils via stored strain energy in a controlled manner using a lanyard and a damper which governs the deployment speed. Covering the boom is a thermal sock which limits temperature excursions and thus diurnal changes in overall length. Atop the boom is a tip/tilt/rotate mechanism (TTR) that may be used on orbit to adjust the initial alignment of the MMAs with respect to the DUs.

**Table 1:** Spacecraft Properties.

| Parameter | Performance |
|---|---|
| **Launch Mass** | 330 kg |
| **Orbital Average Power** | 306 W (End of Life) |
| **Nominal Lifetime** | 2 years (no life-limiting consumables) |
| **Stabilization Method** | 3-axis |
| **Pointing Modes** | Acquire Sun (Safe Mode)<br>Point (Operations Mode) |
| **Attitude Control** | 40 arcsec ($3\sigma$) x & y<br>(In Operations Mode) |
| **Bus Voltage** | 30±4 Volts |
| **Communication Frequency** | S-Band |
| **Command Rate** | 2 kbs Uplink |
| **Telemetry Rate** | 2 Mbps Downlink |
| **Storage** | 6-Gbytes |

## 6. THE GROUND SYSTEM

A diagram showing the entire ground system including communication links is shown is Figure 5. Communication with IXPE is via a primary ground station in Malindi, Kenya, with a backup station in Singapore. Data download will vary with observing program but will average around 6-7 contacts per day. Commanding will be approximately once per week. NASA's Tracking and Data Relay Satellite (TDRS) will be available during the launch and early operations phase of the mission.

Mission operations will be run from the MOC. The MOC communicates with the SOC and with the observatory via the ground network shown in Figure 5.

The SOC is responsible for science operations. The SOC formulates the observing plan that is then sent to the MOC for detailed scheduling. The SOC receives all Observatory data which it then processes. Science data products are then distributed to the HEASARC for public access within 1 week after the end of an observation (after an initial 3-month checkout period during which the turnaround time is 30 days).

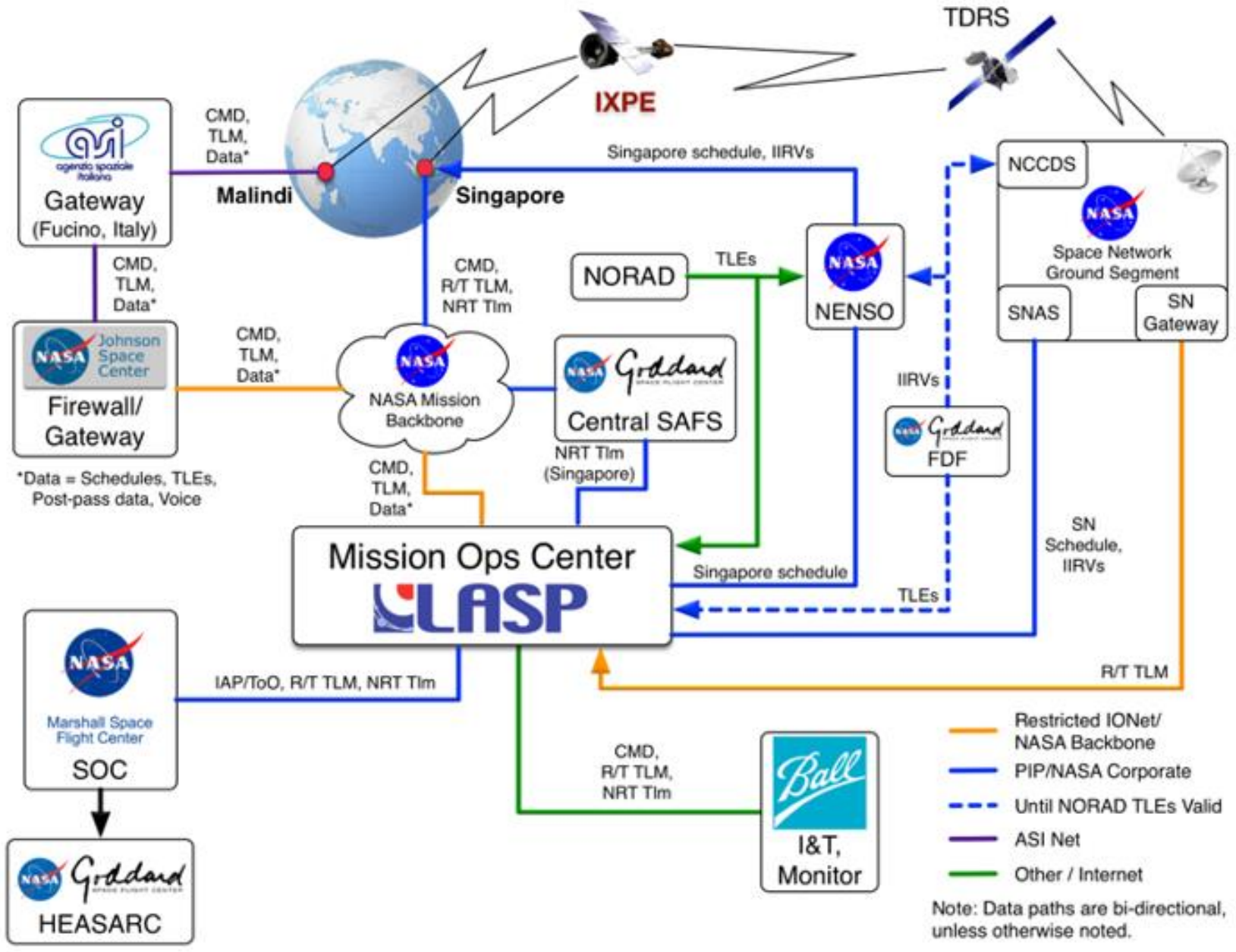

**Figure 5:** The IXPE ground system. The TDRS is only baselined for use during launch and early operations phases of the mission.

## 7. THE SCIENCE PAYLOAD

### 7.1. Mirror Module Assembly (MMA)

The IXPE Project built four MMAs [6,7] each comprised of 24 concentrically nested mirror shells. Three of the MMAs are for flight, the fourth is a spare. The mirror shells are fabricated by electroforming using a nickel/cobalt alloy that has a higher strength than more-typically-used pure nickel. The shells are closely packed, with about a 2 mm separation, to maximize the effective area for a given outer diameter. Each shell has a length of 600 mm which includes both parabolic and hyperbolic segments of the Wolter-1 configuration replicated from the mandrel. There is no additional coating on the inside of the mirror shells, the nickel/cobalt providing optimum reflectivity over the IXPE energy band of 2–8 keV.

The mechanical design of the MMA is shown in Figure 6. The front spider is the primary structural element through which the MMA mounts to the mirror module support structure. Each of the nine spokes of the front spider has a precisely fabricated and positioned comb glued to it. During alignment and assembly, each mirror shell is inserted and glued into the corresponding slot between the tines of each comb, thus mounting the mirror shells to the spokes of the front spider.

The rear spider has 18 spokes, each also with a metal comb. After the 24 shells have been epoxied to the front spider, the rear spider is attached to the central support tube and its metal combs aligned such that each mirror shell floats within the corresponding slot between the tines of the rear-spider combs. However, unlike the front-spider combs which hold the mirror shells, the rear-spider combs merely limit excursions of the shells under launch loads to preclude shell-to-shell collisions. In this way the mirror shells are not over-constrained, as they would be if rigidly attached at both ends. To prevent any possible marring of the mirror shell surface, the rear comb tines have heat-shrink sleeving applied to cushion impacts.

Completing the MMA assembly is a pair of thermal shields, one at each end and provided by Nagoya University in Japan. These shields are fabricated from an ultra-thin (1.4 µm) polyimide film, coated with 50 nm of aluminum and supported on a highly transparent steel mesh. The shields provide for thermal control, restricting heat loss while allowing X-rays to pass though. Table 2 gives MMA parameters.

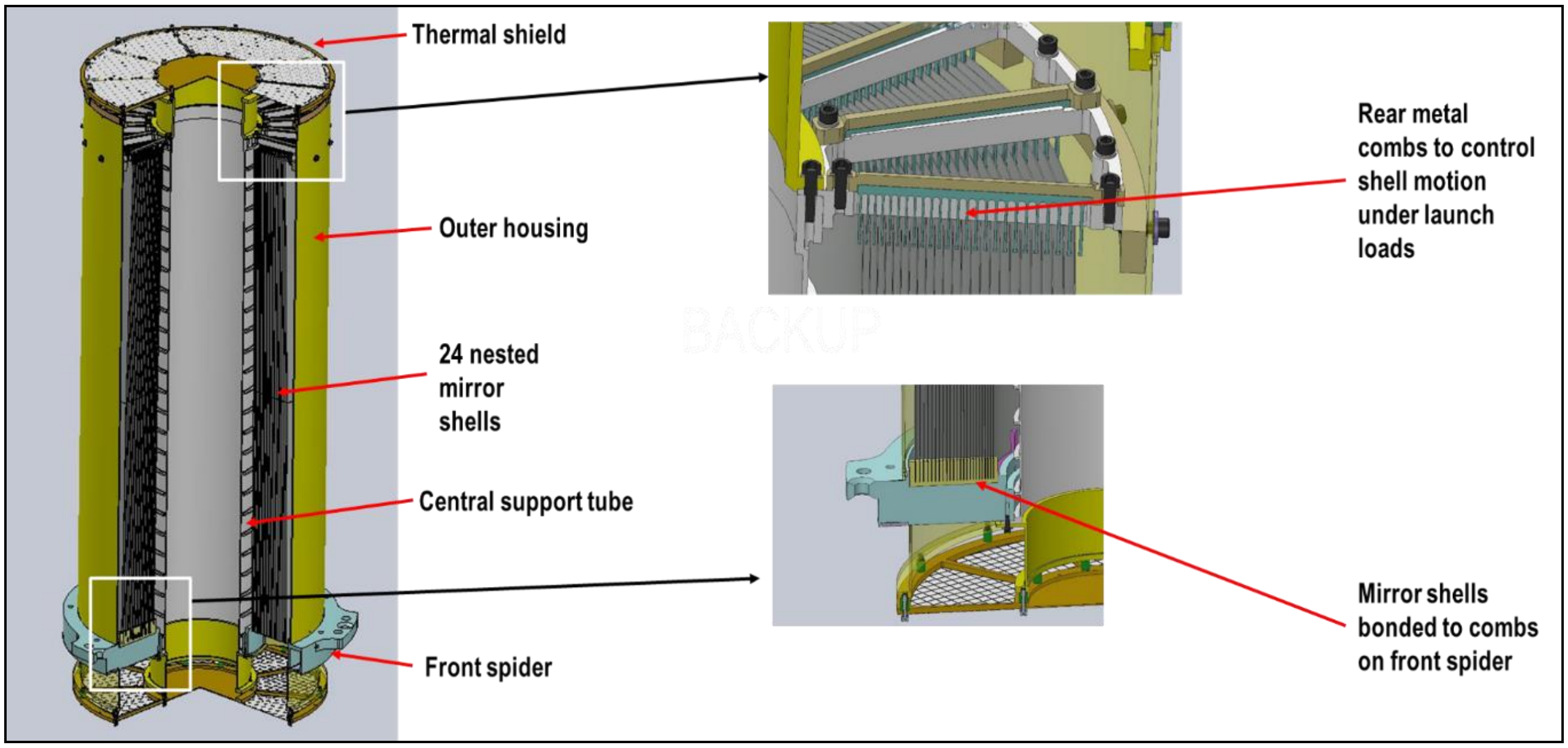

**Figure 6:** Mechanical design of the Mirror Module Assembly

**Table 2:** Mirror Module Assembly Parameters

| Parameter | Value |
| --- | --- |
| Number of mirror modules | 3 |
| Number of shells per mirror module | 24 |
| Focal length | 4 m |
| Total shell length | 600 mm |
| Range of shell diameters | 162–272 mm |
| Range of shell thicknesses | 0.18–0.25 mm |
| Shell material | Electroformed nickel–cobalt alloy |
| **Field of view (detector limited)** | **12.9 arcmin square** |

The telescope field of view (FOV) is limited by the size of the detector to 12.9 ′ × 12.9′ (due to the full detector area of 15 × 15 mm). The response is energy dependent. At low energies, (~2 keV), the response drops to 73 % of its on-axis value at the edge of the FOV (6.45′ off-axis), whereas at 8 keV it drops to just 40% of its on-axis value.

#### 7.1.1. MMA X-Ray Calibration

Precise calibration of the X-ray telescopes is very important for subsequent data analysis. This is particularly true for a pathfinder mission like IXPE where, except for the polarization of the integrated emission from the Crab Nebula at 2.6 keV, measured with the crystal polarimeter aboard the OSO-8 satellite [8], there is no standard to compare to. In fact, due to the dynamical nature of the Nebula and the rotation of the pulsar, even this measurement does not serve as an absolute standard. MMA calibrations were carried out at the MSFC Stray Light Test Facility (SLTF), which features a 100 m vacuum beam tube with X-ray source assemblies at one end and a 10 m × 3 m test chamber at the other (see Figure 7). The Facility is now used mainly for testing X-ray optics, gratings, and detectors. The MSFC X-ray Astronomy Team provides and maintains an array of scientific equipment and personnel to support X-ray testing including X-ray sources, detectors, mechanical stages, and associated support hardware. MSFC also provides personnel to operate the Facility.

MMA calibration measurements were made at multiple energies for on- and off-axis effective area and angular resolution. Also measured were the MMA responses to sources outside of the direct field of view, so called ghost rays, which can increase the background for observations in crowded source regions.

##### 7.1.1.1. The Stray Light Test Facility (SLTF)

The SLTF beamline consists of a 1.2-m-diameter beam tube feeding a 3-m diameter (14 meters long with taper) test chamber. In addition to 3 large roughing pumps able to pump down the beamline to $5\times10^{-3}$ Torr in two hours, there are three large cryopumps that allow the facility to reach a few $\times10^{-6}$ Torr after an additional 2-3 hours. The MMA under calibration is mounted in the test chamber atop a hexapod that enables precise linear control in all axes and precise angular control in tip and pan. At a distance of 4.17 m along the optical axis (the effective focal length of the MMA at 100-m source distance) are the facility detectors used to measure the MMA response. These detectors consisted of a fast silicon drift detector with known effective area and a Charge Coupled Device (CCD) camera with fine (13.5 µm) pixels to measure the MMA point spread function. Both detectors are mounted on a high-precision moveable stage to allow positioning perpendicular to the beam direction. The precision is about 2 um and the stages can carry ~ 36 kilograms over a range of ~600 mm in both X and Y. With this system one can: precisely position the various focal plane detectors in the center of the focused beam from the optic; switch between the various detectors mounted on the detector plate; and move the detectors off to the side to enable direct measurements of the incident X-ray flux unobstructed by the optic under test. After initial MMA calibrations the flight spare detector unit was also mounted on the detector stage.

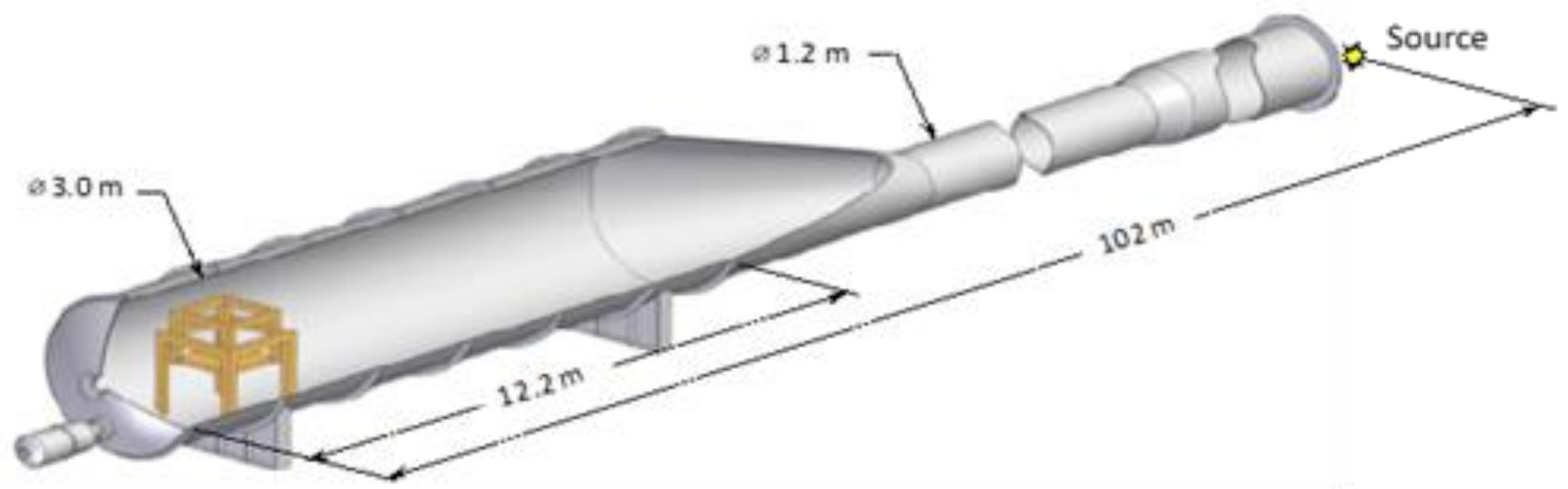

**Figure 7:** The MSFC Stray-Light Test Facility (SLTF)

The hexapod to which an MMA was mounted is vacuum compatible and has a precision better than 0.5 micron in X, Y, and Z, and better than 1 arcsecond in the three rotation angles. The hexapod has a linear range of ± 50 mm in X and Y, and ± 25 mm in Z, and a load capacity of 40 kg. There is also software to allow the hexapod to be "dithered" (moved in a continuous Lissajous-type pattern in X and Y) to simulate spacecraft dithering during flight.

The primary X-ray sources used for X-ray calibration at SLTF are Trufocus models 6050 and 6051 L. The model 6050 is air cooled and has a maximum beam current of 1 mA, while the 6051 is water cooled with a maximum beam current of 2 mA (otherwise, except for the anode material, the sources are identical). IXPE calibration used model 6050 tubes with Mo and Fe anodes, and a model 6051 with a Ti anode. These tubes are vacuum sealed with a 0.005-inch thick Be window, and have a focal spot size smaller than 1 mm. The Facility also has a small Manson source with an Al 6061 anode, capable of operating up to a potential of 10 kV, with a beam current up to about 0.75 mA. This source was used for one IXPE measurement (spurious modulation from an unpolarized source) with an Al filter with optical depth of two.

IXPE also purchased custom X-ray tubes from Trufocus for use with matched crystals as polarized X-ray sources. The characteristics of these polarized sources are shown in Table 3, These sources can operate with potentials up to 30 kV and beam currents nominally up to 30 mA for an X-ray power of ~900 Watts. IXPE used three of these water-cooled windowless sources with Rh, Ti, and Fe anodes coupled with Ge (111), Si (220), and a Si (400) crystals to produce >99% polarized X-rays.

**Table 3:** Characteristics of the SLTF Polarized Sources

| Crystal | X-ray Tube | Energy (keV) | Diffraction Angle | Polarization (%) |
|---|---|---|---|---|
| Ge (111) | Rh L | 2.70 | 44.9 | ~ 100 |
| Si (220) | Ti K | 4.51 | 45.7 | 99.5 |
| aSi (400) | Fe K | 6.40 | 45.5 | ~ 100 |

There is also an X-ray filter system to reduce the continuum emission and emphasize the characteristic X-ray line to produce a more monochromatic source. Filters include Al (for use with Al-K), Nb (for use with Mo-L), Pd (for use with Rh), and Ti, Fe, and Ni (for use with Cu). The filter thicknesses all have an optical depth of 2. There is also an optical blocking filter for use with the windowless X-ray sources. This filter is aluminized polyimide with a circular aperture of 1cm and mounted in a CF 2.75 gate valve. In addition to serving as a filter to block optical light emitted from the X-ray source, the valve also served as a vacuum separation gate to keep the pressure in the source system at $<10^{-6}$ Torr to protect the windowless sources from possible pressure variations in the rest of the beamline.

#### 7.1.2. MMA Calibration Results

During calibration of the MMAs at the SLTF, X-ray sources of characteristic energy of 2.3 keV (Molybdenum L) or 4.5 keV (Titanium K) illuminate the optic, the flux from which is then focused onto a silicon drift detector (SDD). A simultaneous measurement of the flux incident onto the front of the optic is made with a separate SDD, the same type as the focal plane SDD and cross-calibrated with it. This measurement approach gives both a continuum effective area covering the IXPE bandpass, of 2.0-8.0 keV and a measurement of the effective area at the X-ray source line energy.

##### 7.1.2.1. Effective Area

The appropriate ratio of the SDD measurements, multiplied by the 50 $mm^2$ SDD area, gives the effective area of the MMA at a finite-source-distance of 97.950 meters (the distance from the X-ray source to the MMA node). A correction, verified by a ray-trace analysis, is applied to the finite-source-distance area to give the on-orbit infinite-source-distance effective area. The on-axis MMA line-source effective areas measured during calibration are given in Table 4. The uncertainty in these are ± 3%, (1-σ). The results of the continuum effective area measurements for each of the three flight MMAs are shown in Figure 8.

**Table 4:** IXPE MMA effective areas as measured at 2.3 and 4.5 keV during calibration at the MSFC SLTF. The areas have been corrected to infinite source distance as discussed in the text.

| | 2.3 keV | 4.5 keV |
|---|---|---|
| **MMA1** | 168 $cm^2$ | 195 $cm^2$ |
| **MMA2** | 167 $cm^2$ | 195 $cm^2$ |
| **MMA3** | 167 $cm^2$ | 200 $cm^2$ |

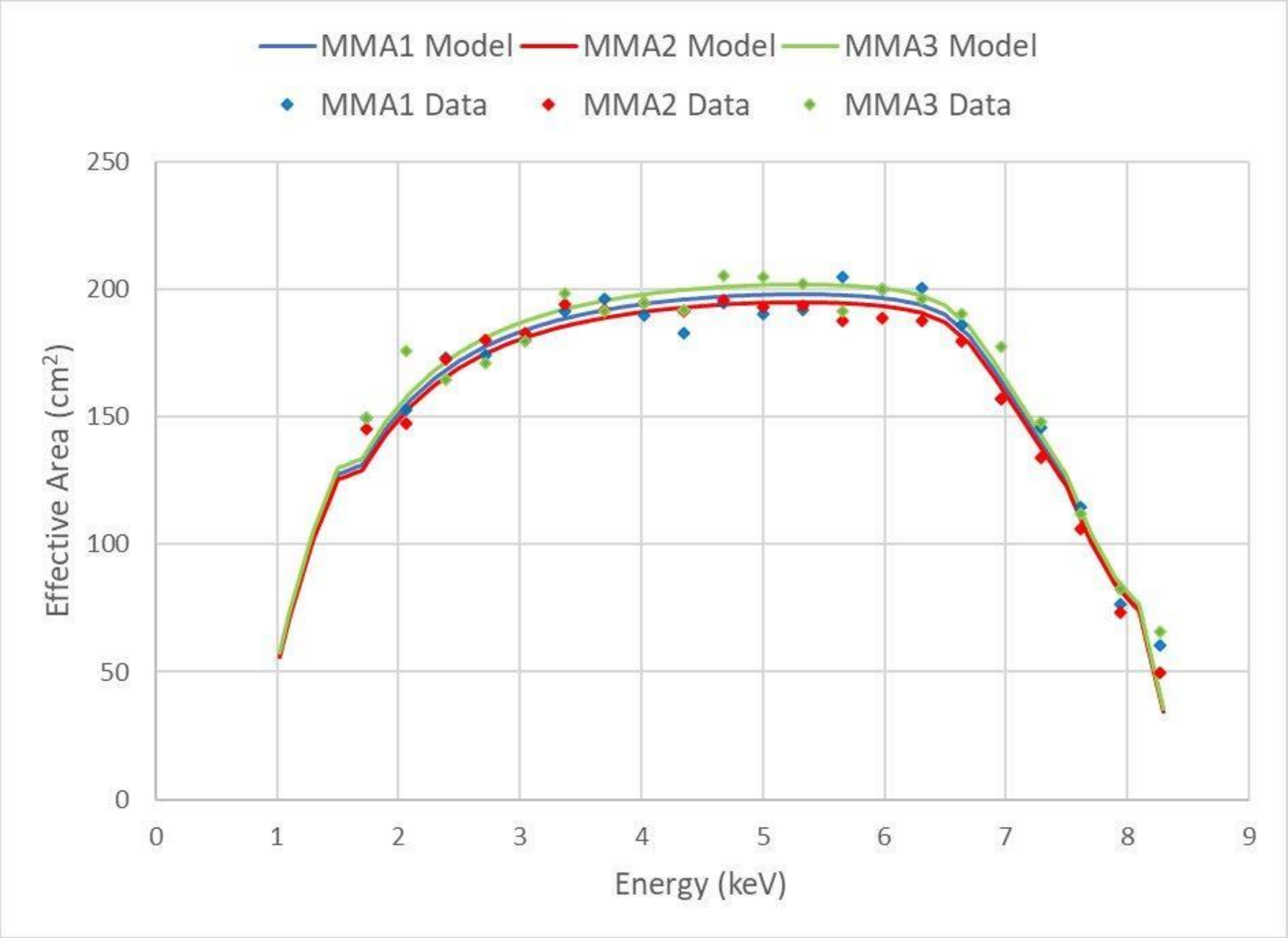

**Figure 8:** Effective area (measured and best-fit model) for a point source at infinity as a function of energy for the three flight MMAs.

### 7.1.2.2. Half Power Diameter

The Half-Power Diameter (HPD) of each MMA was measured using a CCD camera at the focus of the optic, with filtered X-ray sources producing lines at nominally 2.3 keV (Mo-L) and 4.5 keV (Ti-K) along with some (filtered) continuum emission. The detector region used for this measurement was chosen to be the same as that used for the effective area, i.e. a circle of diameter 8 mm, which is approximately 16 times the MMA HPD. Flat fields, taken immediately before and after the measurements, were used to subtract offsets and noise contributions in individual pixels. The diameter containing half the flux within the measurement region was then determined and converted to an angle using the measured image distance (~ 4.17m) for the 100-m object distance at the SLTF. Results are tabulated in Table 5.

**Table 5:** Measured Half-Power Diameter at 2.3 keV and 4.5 keV for MMA1, 2, and 3.

| | MMA1 | | MMA2 | | MMA3 | |
|---|---|---|---|---|---|---|
| **Energy (keV)** | 2.3 | 4.5 | 2.3 | 4.5 | 2.3 | 4.5 |
| **HPD (arcsec)** | 19 | 19.9 | 25 | 26 | 27.6 | 28 |

## 7.2. The Detector Units

The heart of the IXPE payload is the DUs. Located at the focus of each MMA, these provide position determination, energy determination, timing information and, most importantly, polarization sensitivity. Inside each DU is a Gas Pixel Detector (GPD) which images the photoelectron tracks produced by X rays absorbed in the special fill gas (Dimethyl Ether - DME). The initial emission direction of the photoelectron determines the polarization of the source, while the initial interaction point and the total charge in the track provide the location and energy of the absorbed X ray, respectively.

Figure 9 shows a schematic of the GPD. An X ray enters through a beryllium window and interacts in the DME fill gas. The resulting photoelectron produces a trail of ionization in the gas, and this photoelectron "track" drifts through a GEM to provide charge gain, and then onto a pixel anode readout. Table 6 gives the relevant DU parameters.

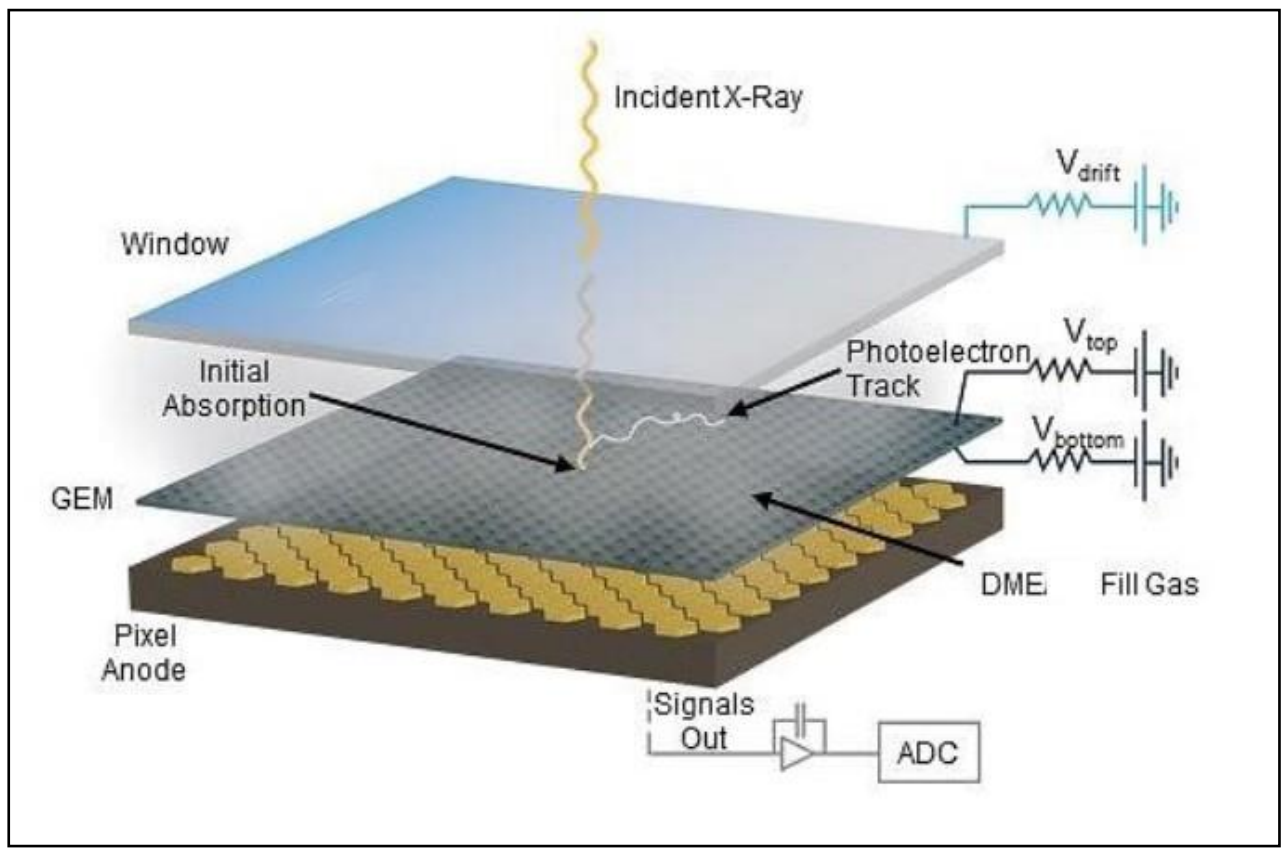

**Figure 9:** Schematic of the Gas Pixel Detector (GPD)

**Table 6:** Parameters of an IXPE Detector Unit (DU)

| Parameter | Value |
|---|---|
| Sensitive area | 15 mm × 15 mm (13 x 13 arcmin) |
| ASIC Readout Pitch | 50 µm |
| Fill gas and asymptotic pressure | DME @ ~ 640 mbar |
| Detector window | 50-µm-thick beryllium |
| Absorption and drift region depth | 10 mm |

| Spatial resolution (FWHM) | ≤ 124 µm (6.4 arcsec) @ 2 keV |
|---|---|
| Energy resolution (FWHM) | 0.52 keV @ 2 keV (∝ $\sqrt{E}$) |
| Useful energy range | 2 - 8 keV |

An expanded view of a detector unit is shown in Figure 10 (left). As well as the GPD, the unit houses the entire "back-end" electronics to process each event, as well as high-voltage power supplies. It also houses a filter and calibration wheel assembly for on-orbit calibration (see section 7.5), as well as a collimator for reduction of X-ray background. At the base of the collimator is a very thin polyimide UV ion shield (not shown) which prevents ions and UV from entering the GPD region. Figure 10 (right) shows a photograph of a completed flight DU.

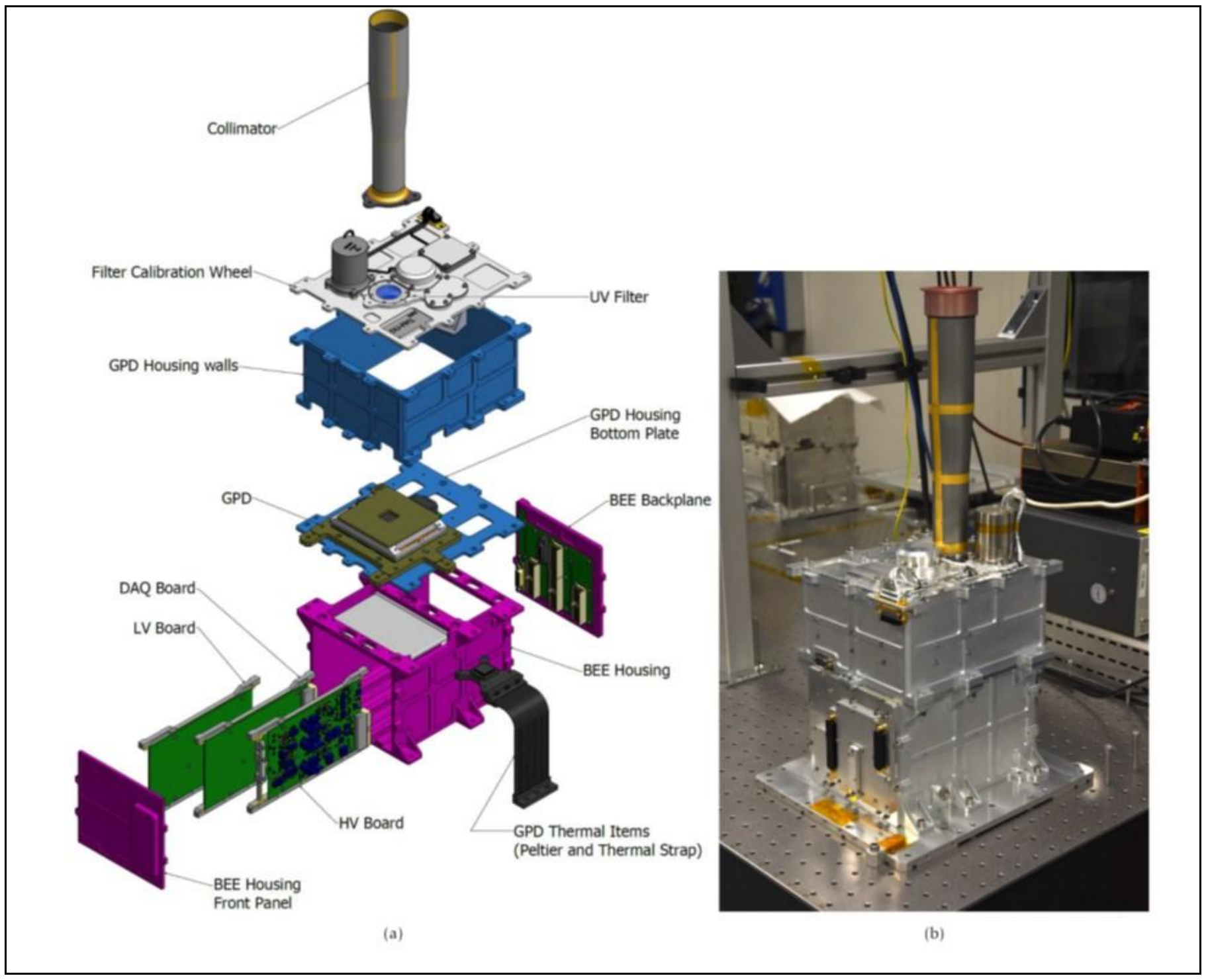

**Figure 10:** Expanded view of a Detector Unit (DU, left); photograph of a completed flight unit (right).

Figure 11 shows a photoelectron track from a 5.9 keV photon, as imaged by the GPD. The image reconstruction starts with a subtraction of a noise floor, followed by a clustering stage which groups together all contiguous pixels above the threshold to highlight the track. Once the physical photoelectron track is separated from the noise pixels, a first moment analysis of the two-dimensional charge distribution is performed, yielding estimates for the barycenter and the principal axis. The sum of the charge content over all the track pixels provides the energy measurement, while the charge asymmetry in the longitudinal profile, due to the Bragg peak, can be exploited to identify the initial part of the track, which is the one carrying most of the information about the original photon polarization direction. At this point, a second moment analysis is run, de-weighting the pixels close to the end of the track, to get a more accurate estimate of the photon absorption point and the photoelectron emission direction.

In the future we will be improving the sensitivity to polarization by applying an advanced technique developed as part of the IXPE project at Stanford University. The approach is based on machine learning and Neural Networks.[9]

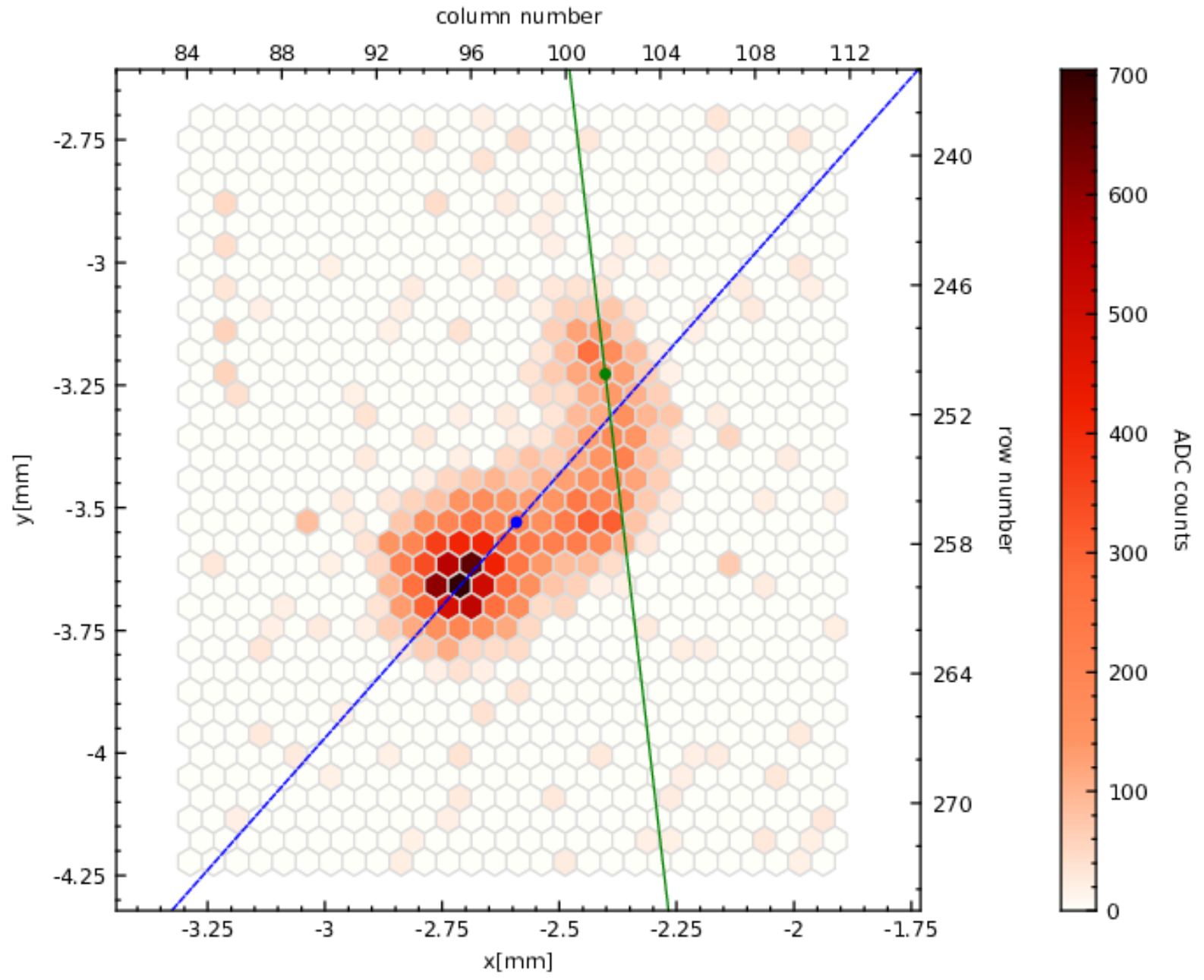

**Figure 11:** Example of the image of a photoelectron induced track from a 5.9 keV photon. [10] The color scale represents the charge content of each pixel after subtracting a noise floor. The blue line and point represent the principal axis and the barycenter of the track, while the green line and point represent the best estimate of the photoelectron direction and photon absorption point using the moment analysis.

#### 7.2.1. DU Calibration

Once constructed and environmentally tested, the DUs and the DSU were ready for X-ray calibration (DUs integrated with the DSU are referred to as the Instrument [11].) Each DU, including both the flight and spare units, went through a comprehensive calibration to characterize the response to both polarized and unpolarized radiation and to measure the spectral, spatial and, timing performance. The DUs were also integrated to the DSU and illuminated with X-ray sources to test the operation of the Instrument in the flight configuration. These activities required equipment to generate X-rays with known polarization and position angle. Calibration of the Instrument relied on custom calibration sources specifically designed and built in Italy for this purpose.

Calibration of flight units started with DU1 on September 6th, 2019, continued with DU2 and ended with DU3 on February 3rd, 2020. Including the flight-spare unit, 530 measurements were performed over a total of 4052.3 hours and involved 2.250 billion counts.

##### 7.2.1.1. Instrument Calibration Equipment (ICE) and Assembly Integration and Verification Test Calibration Equipment (ACE)

There are two Instrument calibration stations, the ICE and ACE. The two stations are shown in Figure 12 and are discussed in detail in the paper by Muleri, et al. [12]. X-ray sources can be mounted on both test stations. However, the ICE offers a full-fledged set of manual and motorized stages to align and move the source beam in a controlled way, even during a measurement, whereas the ACE was equipped only with a subset of available sources. Nevertheless, these sources provided the capability to perform tests with ACE and, contemporaneously, with ICE. The most time-consuming calibrations are the calibration of the response to unpolarized radiation because of statistical considerations.

A set of manual and motorized stages allow adjustment of the beam direction with respect to the unit in calibration. These stages allow: movement of the DU in the plane orthogonal to the incident beam with an accuracy of 2 μm (over a range of 100 mm); to rotation of the DU on the azimuthal plane orthogonal to the incident beam with an accuracy of 7 arcsec to measure the response at different polarization angles; tip/tilting of the DU to align it to the incident beam and; averaging residual polarization of unpolarized sources. The X-ray source is mounted on a mechanical support that permits adjustment of its position and inclination with respect to the DU. Finally, a manual translation stage allows one to slide the source assembly to three separate positions to perform the calibration with the DU or to illuminate commercial detectors.

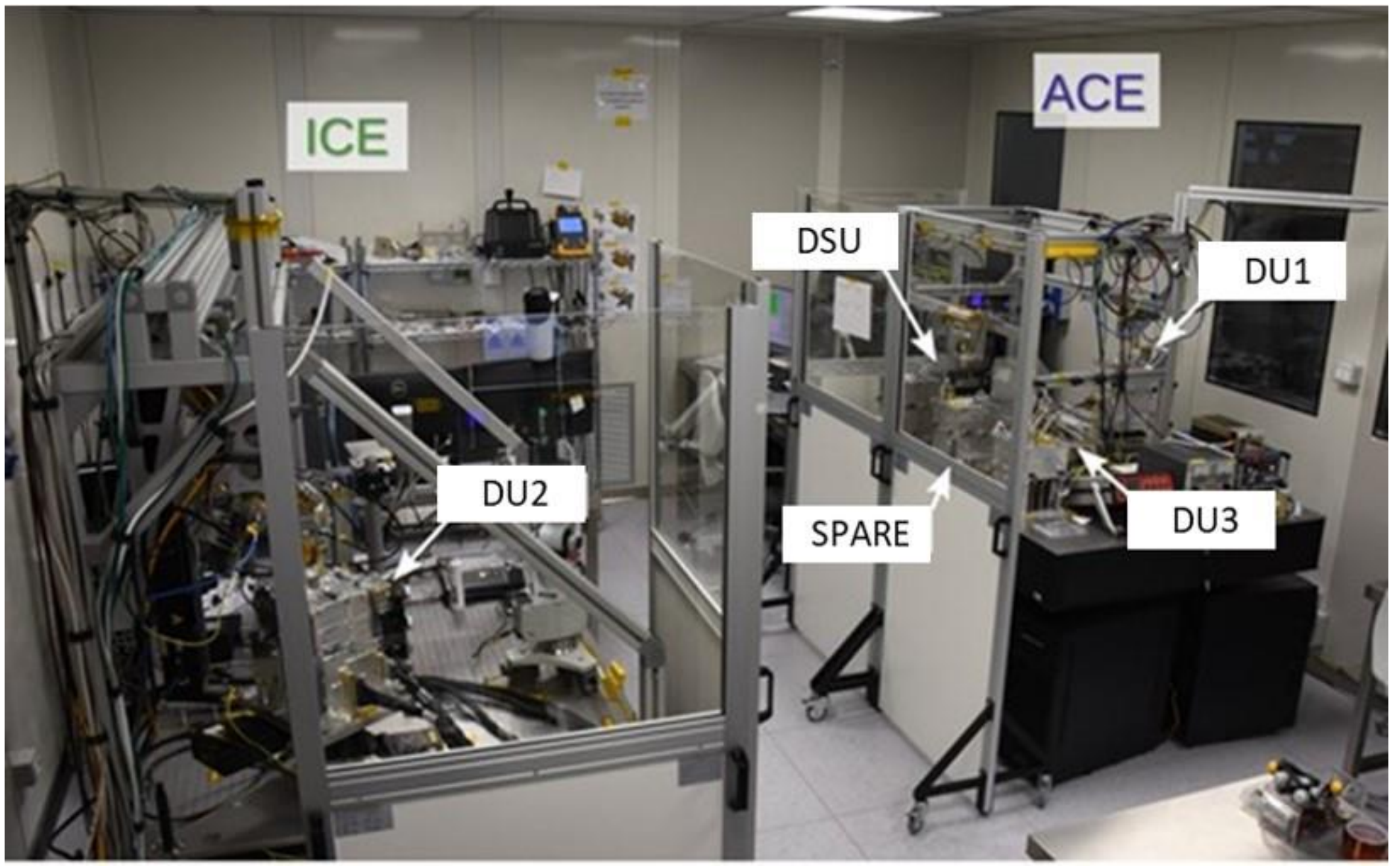

**Figure 12:** The ICE and ACE calibration stations in Italy during calibration of DU2 flight model on the ICE and the Instrument testing with DSU-EM (Engineering Model) and DU1,3 (flight models) and the spare unit on the ACE.

#### 7.2.1.2. Unpolarized Sources

Unpolarized sources in the ICE cover the entire 2-8 keV energy range of IXPE (Table 7). Only in a few cases it is possible to use a "genuine" unpolarized source where any intrinsic source polarization is much lower than the statistical significance achieved in the calibration. To ensure that the unpolarized sources are truly unpolarized we adopted a simple procedure to decouple any intrinsic source polarization and the response of the instrument to completely unpolarized radiation. The procedure, described in Rankin et.al [13], is based on the repetition of the measurement with the source set at two different azimuthal angles differing by 90°.

Both radioactive sources and X-ray tubes were used. Radioactive nuclei emit unpolarized radiation and $^{55}$Fe sources were used. This nuclide has half-life of 2.737 years and it decays into $^{55}$Mn for K-capture with emission of Mn fluorescence lines K$\alpha$ (5.90 keV, 90%) and K$\beta$ (6.49 keV, 10%). Both lines are expected to be unpolarized from first principles and no hint of polarization was found (see Table 7 which also lists the intrinsic polarization measured from the other X-ray sources).

Measurements were carried out with X-ray tubes in two different configurations: one direct and one for the extraction of fluorescence emission from a target. In the direct configuration, the DU is illuminated with the direct emission of the X-ray tube (or $^{55}$Fe), which is comprised of unpolarized fluorescence lines plus continuum emission. The latter may be partially polarized depending on the details of the X-ray tube emission geometry.

**Table 7:** Intrinsic polarization of the ICE X-ray sources.

| Energy | Configuration | Source Setting | Rate (c/s) | Polarization (%) |
|---|---|---|---|---|
| **2.04** | Fluorescence of Zr target illuminated by Rh X-ray tube | 19 kV, 0.95 mA | ~ 175 | 1.02 % +/- 0.19% |
| **2.29** | Fluorescence of Mo tube illuminated by Ag X-ray tube | 17 kV, 0.77 mA | ~ 190 | 0.72% +/- 0.12% |
| **2.70** | Direct X-ray tube with Rh anode + 300 μm PVC filter | 4 kV, 0.57 mA | ~ 200 | 6.45% +/- 0.1% |
| **2.98** | Direct X-ray tube with Ag anode + 6 μm Ag filter | 4 kV, 0.05 mA | ~ 200 | 12.45% +/-0.09% |
| **3.69** | Direct X-ray tube with Ca anode | 5.4 kV, 0.03 mA | ~ 200 | Undetected MDP (99%) = 0.23% |
| **5.89** | $^{55}$Fe nuclide | 4 mCi | ~ 180 | Undetected MDP (99%) = 0.19% |

### 7.2.1.3. Polarized Sources

The polarized sources exploit the use of Bragg reflection as close to 45 degrees as possible to produce an almost 100%-polarized beam. The characteristics of these sources are listed in Table 8.

**Table 8:** Polarized sources used during DU calibrations

| Crystal | X-ray Tube | Energy (keV) | Diffraction Angle | Polarization (%) |
|---|---|---|---|---|
| PET (002) | Ti | 2.01 | 45.0 | ~ 100 |
| InSb (111) | Mo L | 2.29 | 46.4 | 99.2 |
| Ge (111) | Rh L | 2.70 | 44.9 | ~ 100 |
| Si (111) | Ag L | 2.98 | 41.6 | 95.1 |
| Al (111) | Ca K | 3.69 | 45.9 | 99.4 |
| Si (220) | Ti K | 4.51 | 45.7 | 99.5 |
| Si (400) | Fe K | 6.40 | 45.5 | ~ 100 |

### 7.2.2. DU Calibration Results

#### 7.2.2.1. Quantum Efficiency

The quantum efficiency of the DU is the product of the quantum efficiency of the GPD and the X-ray transmission of the UV/ion filter placed at the bottom of the collimator. It was measured during calibration at INAF-IAPS for each DU. [14] The X-ray transmission of the ion-UV filter was independently measured for all the flight units and is compatible with the nominal transmission expected for the filter composition (1.06 µm of LUXFilm® polyimide + 50 nm Al). [15]

The quantum efficiency of the GPD depends substantially on the absorption gap thickness, the gas pressure, and the transmission of the Beryllium window. The absorption gap thickness is known with high precision by metrology measurements, whereas the gas pressure decreases with time due to the capture of gas molecules by the materials inside the GPD gas cell. The pressure variation has been closely monitored since a few days after initial gas filling and shows a well-established progression to an asymptotic value which will be substantially reached by launch (see section 7.2.2.8). The quantum efficiency expected at launch is shown in Table 9, below.

**Table 9:** DU quantum efficiency assuming a launch date of Dec 9, 2021

| DU# | Energy (keV) | Quantum Efficiency at Launch (%) |
|---|---|---|
| **DU1** | 2.697 | 13.7 |
| | 6.400 | 1.77 |
| **DU2** | 2.697 | 13.4 |
| | 6.400 | 1.72 |
| **DU3** | 2.697 | 13.5 |
| | 6.400 | 1.74 |

Each DU contains a gray (attenuating) filter on the filter and calibration wheel for handling high intensity sources. The X-ray transmission of the gray filter was measured at different energies. [11] The measured transmission is consistent with the nominal thickness of the filter --- a 3 mil (76.2 µm) Kapton foil allowing for tolerances. Results are shown in Table 10.

**Table 10:** Gray filter x-ray transmission.

| DU# | Energy (keV) | Transmission (%) | Transmission 1σ (%) |
|---|---|---|---|
| **DU1** | 2.697 | 17.19 | 0.17 |
| | 6.400 | 87.73 | 0.52 |
| **DU2** | 2.697 | 18.13 | 0.17 |
| | 6.400 | 88.39 | 0.64 |
| **DU3** | 2.697 | 17.01 | 0.16 |
| | 6.400 | 88.04 | 0.78 |

#### 7.2.2.2. Energy Resolution

We made use of flat field data obtained with the polarized sources to measure the energy resolution. Because of the use of Bragg crystals these sources are highly monochromatic. Spectral data were taken in 100 spots of 500-µm radius from the flat field data. The energy resolution was then calculated by fitting with a Gaussian the gain-corrected energy spectrum and dividing the measured FWHM by its peak. The energy resolution as a function of energy for the three flight DUs is shown in Figure 13.

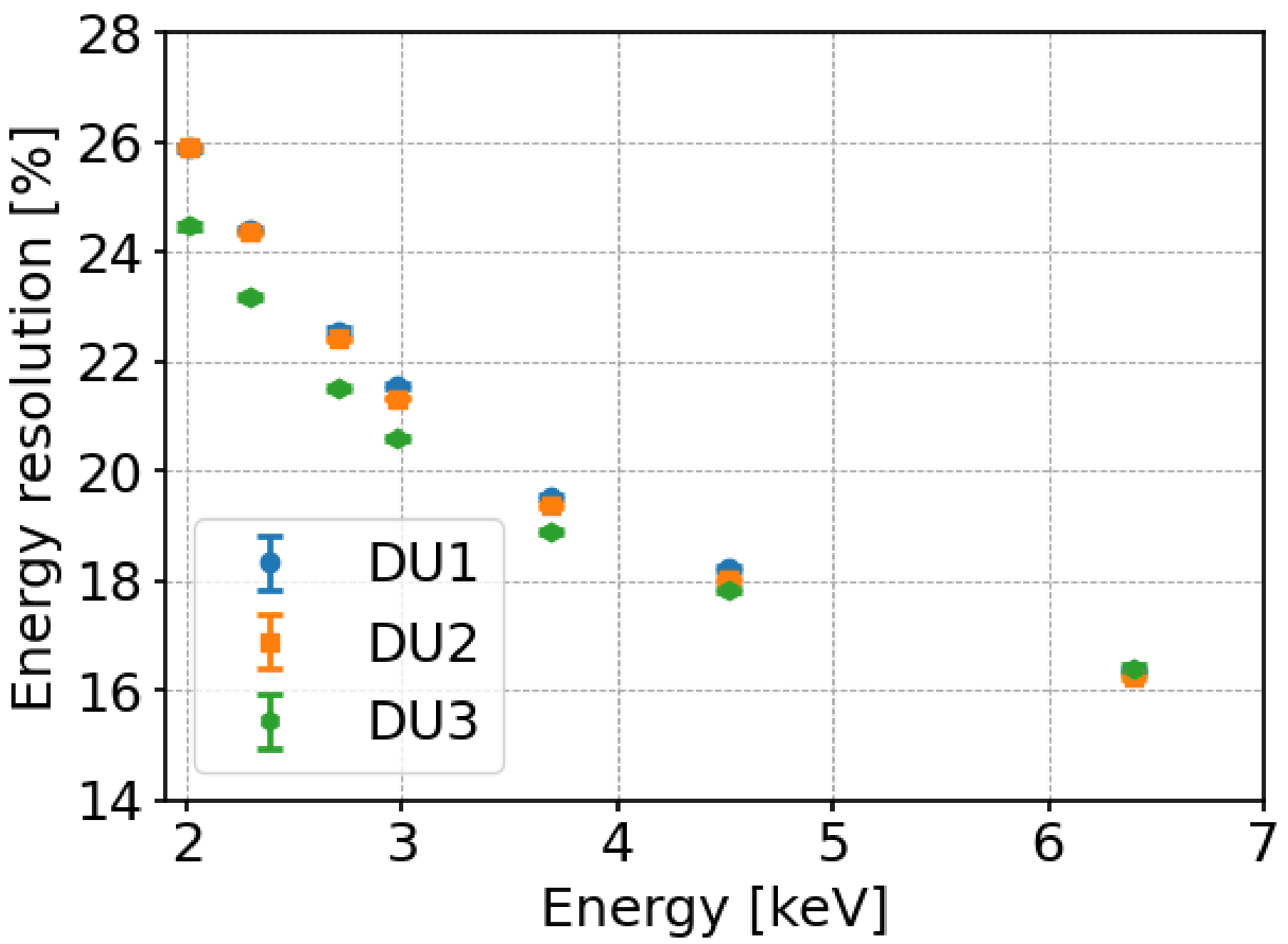

**Figure 13:** Energy resolution as a function of energy for the flight DUs.

#### 7.2.2.3. Timing

The GPD provides, for each processed event, a trigger output which is used to time tag each event with 1 µs resolution and 1-2 us accuracy thanks to the use of a Global Positioning System (GPS) that provides a pulse per second (PPS), and temperature-controlled 1-MHz Oscillators. The time for ASIC readout and subsequent processing by the back-end electronics is such that the dead time is 1.1-1.2 ms (depending on energy) per event. This gives about 12 % deadtime for the bright Crab Nebula. Note that the polarization response of the ASIC does not depend on the dead time fraction (e.g. on the source rate).

#### 7.2.2.4. Spatial Resolution

During ground calibrations the intrinsic spatial resolution was estimated by measuring the HPD using an on-axis beam with a diameter less than 39 µm (measured by a CCD test detector). Measurements were performed for both polarized and unpolarized sources and in different locations across the sensitive area of the detectors. Details of these measurements are given in a dedicated paper [16]. Table 9 shows the average of results obtained for the 3 flight detector units.

**Table 9:** Mean spatial resolution of IXPE flight DUs as a function of energy.

| Energy (keV) | HPD (µm) |
|---|---|
| **2.70** | **116.28 +/- 0.09** |
| **6.40** | **123.5 +/- 0.4** |

### 7.2.2.5. Modulation Factor

The modulation amplitude is the fractional variation in the number of detected events versus position angle. The "modulation factor" is the modulation amplitude for 100% linearly polarized radiation. The calibration of the modulation factor is described in detail in the paper by Di Marco et al. [16]. The modulation factor was measured at different energies and angles to estimate the accuracy of the polarization angle and to check that spurious modulation (see section 7.2.2.6) is properly subtracted. An important outcome of these measurements is that the modulation factor of the flight detectors is uniform over the detector sensitive surface and is not dependent on the position angle.

Calibration data have been analyzed subtracting spurious modulation Stokes parameters using the methods described in Rankin et al., [13] and using the weighting approach described in Di Marco et al.[17] From these data, and accounting for the secular change in pressure (see section 7.2.2.8), it is possible to estimate the modulation factor as a function of energy as shown in Figure 14 for the time of launch set to 9 December 2021.

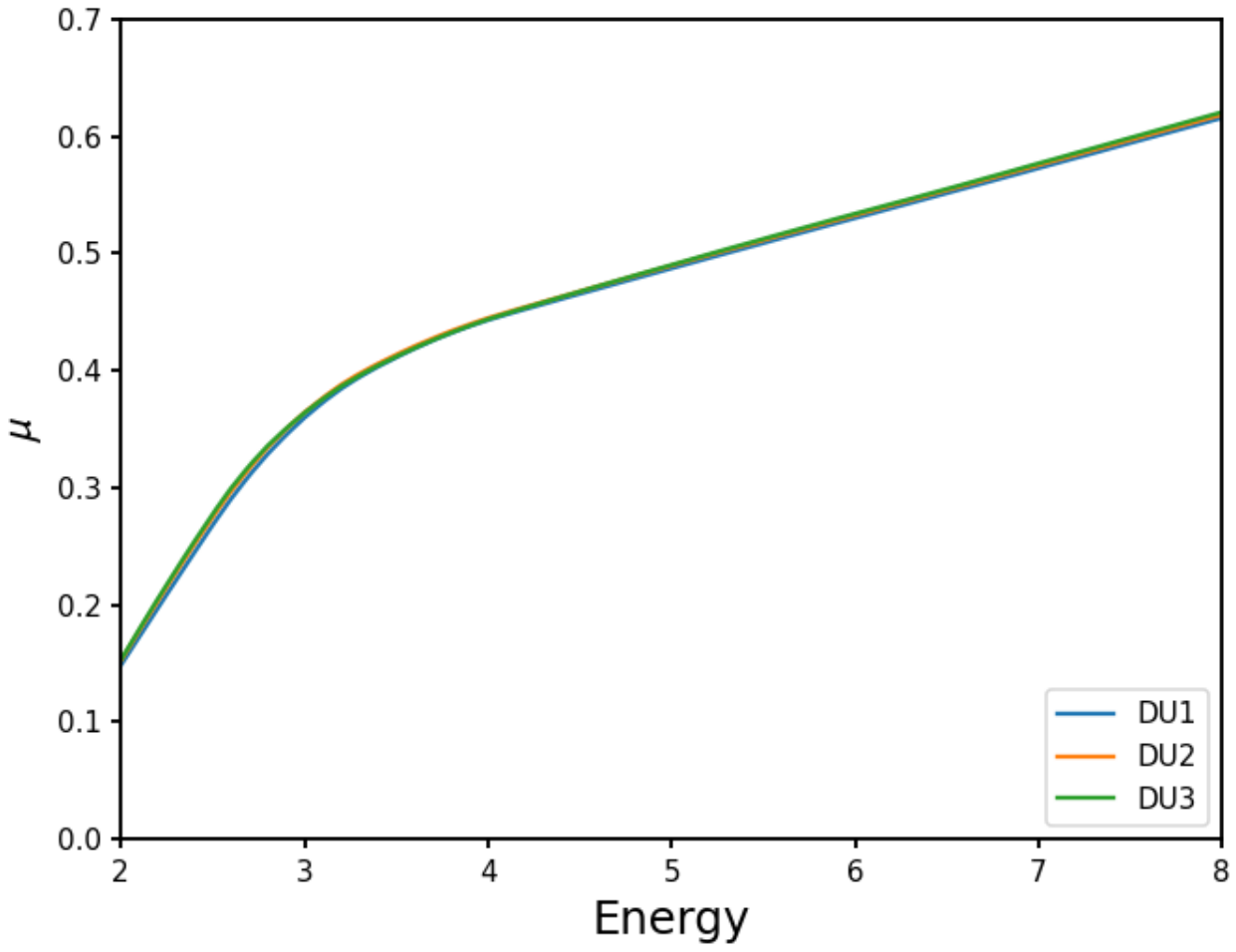

**Figure 14:** Modulation factor as function of energy for the IXPE DUs estimated at time of IXPE launch. Values are derived from ground calibration data after correcting for spurious modulation and applying weighting.

### 7.2.2.6. Spurious modulation

The response of an ideal polarimeter to unpolarized radiation would show no variation of the position angle as the device is rotated about the line of sight. However, the IXPE focal plane detectors show, especially at low energy, a low amplitude, nearly-$\cos^2$ variation of the position angle for unpolarized X-rays in detector coordinates. Although the root causes of such spurious modulation have been thoroughly investigated [10], the effect is hard to model and therefore was fully calibrated. The response of each DU, including the flight spare, was measured at six energies over the entire field of view, and with higher statistics in its center where the brightest sources will be observed.

The Stokes parameters of the spurious modulation are shown as a function of energy in Figure 15. In the plot, the contribution from the three IXPE DUs are appropriately summed accounting for the fact that they are clocked at 120° with respect to each other. Spurious modulation has also been shown to be constant with temperature and counting rate by tests conducted with a number of prototype GPDs. The spurious modulation will be removed, event-by-event, during data processing [11].

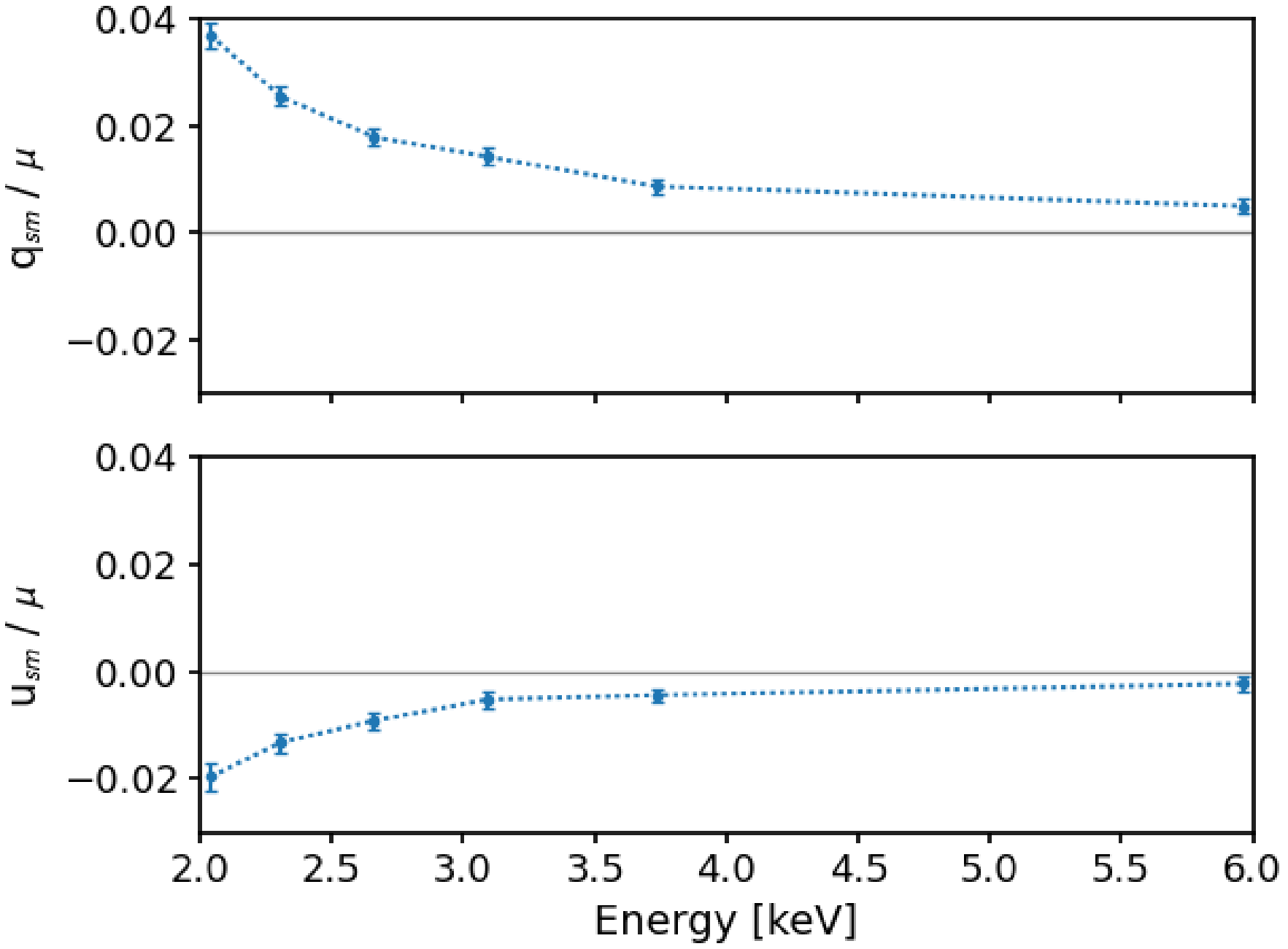

**Figure 25:** Stokes parameters as a function of energy of the detector-averaged spurious modulation accounting for the 120° clocking of the detectors. The effect is removed during the on-ground data processing.

#### 7.2.2.7. Rate-dependent gain variations

The fine pitch of the IXPE GEMs makes them susceptible to rate-dependent gain variations due to temporary charge build-up. When the detector is irradiated, part of the charge from the avalanche can be temporarily deposited onto the dielectric substrate in the GEM holes, affecting the configuration of the electric field, and causing local (and reversible) changes in the gas gain. Since the charge trapping is not permanent, a competing discharging process is continuously at play, causing the gain to drift toward the initial value when the input energy flux is low enough.

Although the charging effect was largely mitigated by a dedicated fine tuning of the production process in the development stage of the mission, it is still present at a level that needs to be corrected for to achieve an accurate estimate of the energy and, since the modulation factor is energy-dependent, to correctly infer the source polarization from the measured modulation. For a typical X-ray celestial source, the charging effect will be of the order of a few %, on timescales of several hours to a day. For very bright sources, we anticipate gain variations up to 10% on much shorter timescales.

To this end, we have developed a complete phenomenological model that allows us to measure and correct such gain variations, based on the energy flux measured by the GPD as a function of time and position across the active surface. (We emphasize that all the necessary information is included in the science data that are collected during normal observations, and that the onboard calibration sources can be used to ensure that there are no systematic drifts over long observations). The reader is referred to section 7.2 and appendix A of this reference [10].

#### 7.2.2.8. Long-term pressure variation

During the development of the IXPE instrument we realized that the internal gas pressure underwent a long-term decrease, over timescales of months, reaching an asymptotic value ~150 mbar below the initial pressure of nominally 800 mbar. This observation, initially hinted at by a slow increase of the gas gain, has since been confirmed by several indirect measurements including the evolution of the quantum efficiency and the average track length at a given energy, as well as direct measurements of the vertical displacement of the Be window.

The effect has been systematically monitored using a dozen (nominally identical) detectors, with test data amounting to more than 30 GPD-years equivalent of operation. We note that the energy resolution shows no sign of degradation with time, clearly indicating that the root cause is not a real leak between the gas cell and the external world. The "virtual leak" is most likely connected to an adsorption phenomenon within the GPD itself. The pressure decrease is unambiguously saturating with time, and in fact, all three IXPE detectors are all already within a few % of the expected asymptotic value (Table 10). Given the convexity of the time evolution, the current time derivative of the pressure, typically of the order of a 1-10 mbar per year, can be taken as a generous upper limit to the additional variations that we might expect on orbit. Also, the onboard calibration sources will allow us to monitor the detector evolution continuously during the mission, through the measurement of the counting rate and the average track length.

**Table 10:** Detector launch and asymptotic pressures

| Detector | Pressure at Launch | Asymptotic Pressure |
|---|---|---|
| **DU1** | 652 | 645 |
| **DU2** | 632 | 631 |
| **DU3** | 639 | 638 |

Calibration data and simulations indicate that the impact of this pressure variation on the polarimetric sensitivity is mild. While the detector quantum efficiency scales linearly with the pressure, causing a net loss of telescope effective area, the modulation factor increases as the pressure decreases, owing to fact that the tracks become more elongated. The net effect of these two competing processes is that the relative loss of sensitivity, expressed as the broadband minimum detectable polarization for a typical source spectrum, is less than 2%. Once the calibrated corrections are made there is no net effect.

#### 7.2.2.9. Background

The IXPE background is expected to be negligible for any planned observations of point-like sources. Polarimetry is in fact photon starved and typically possible only with the brighter sources for each class, and this reduces the background requirement on the detectors. However, for some observations of extended and low-brightness sources (such as molecular clouds in the Galactic center and SN1006) background may be an issue and some background suppression techniques need to be applied to the data. We evaluated the background expected for IXPE [18] by modeling the spacecraft, detailing the detector geometry, and considering the major components of background such as Cosmic X-ray background, albedo Gamma-ray and albedo neutrons, primary electrons, positrons, protons and alphas, and secondary electrons, positrons and neutrons. The expected counting rate without any background rejection, other than the energy threshold (2-8 keV) is $4.7\times10^{-2}$ c/s/cm$^2$/DU. After background rejection techniques are applied this becomes $1.16\times10^{-3}$ c/s/cm$^2$/DU. Background rejection includes track size, track skewness, track elongation, charge density, cluster number and border pixels The major contribution of un-rejected background are delta-rays from electron and positrons which produce tracks indistinguishable from photoelectron tracks.

### 7.3. Telescope Calibration results

In previous sections we described the independent X-ray calibration of the DUs and the MMAs. A cross-check on these analytic calculations was performed by X-ray-calibrating the spare MMA with the spare DU at the SLTF. Angular resolution (half-power diameter) and effective area data were collected as the MMA was stepped through on-axis and off-axis angles. As rotating the optic about its node does not move the focal image, the detector was also stepped to the appropriate off-axis positions. As part of the calibration, the telescope modulation factor was also measured. This was done while dithering the image on the detector as will be done in flight. The measurement was important to show that the addition of the MMA would have no effect on the modulation factors already measured during detector calibration in Italy.

### 7.3.1. Effective area

The MMA effective area folded with the detector efficiency is a principal input to the IXPE sensitivity calculation - the other driver is the modulation factor. All three of these factors are functions of energy. Effective area data were taken with Mo L, Ti K, and Fe K characteristic X-rays. To suppress continuum emission, each source was filtered using Nb (5 µm), Ti (77 µm) and Fe (77 µm) respectively. A monitor counter, positioned at the MMA, was used to derive the incident flux. This device, a fast Silicon Drift Detector (SDD) has known collecting area and quantum efficiency. For each energy and angle, a minimum of 1000 counts were collected in the monitor counter. As the energy resolution of the DU is about 17% FWHM at 6.4 keV, it cannot resolve the $K_\alpha$ from the $K_\beta$ lines for either Ti or Fe, or the different L lines from molybdenum; therefore, a weighted mean energy was calculated for these measurements. The data from the monitor SDD, which has high spectral resolution, is the sum of all the K lines for each source. The L lines from molybdenum are not resolved by the SDD. The effective weighted energies for the three x-ray sources, correcting for SDD efficiency, are: 2.3 keV for Mo L, 4.6 keV for Ti K and 6.6 keV for Fe K.

The data were processed using a standard analysis based on the moments of the photo-electron-induced tracks (see section 7.2). Because the DU has limited energy resolution and there are low-energy events present due to incomplete charge collection, the acceptance band of the DU was opened to + 3σ above the peak energy and everything below the peak. Similarly, the acceptance band of the SDD detector was opened to include all $K_\alpha$ and $K_\beta$ lines plus any escape peaks and low-energy events. The DU data was deadtime-corrected, and the resulting count rate is divided by the rate from the SDD flux monitor and multiplied by the flux-monitor area. Finally, the resulting effective areas were corrected to that for infinite source distance, using appropriate energy and off-axis-angle dependent correction factors. The resulting effective areas (Area 1) are shown in Table 11 where they are compared to the analytic combination (Area 2) for the DU and MMA calibrated separately. The Area-2 measurements were corrected for the known pressure drop that took place between the time of these measurements. Predicted and measured telescope effective areas are within statistical errors, as expected. The same process was repeated, at each energy, for several off-axis angles with similar results.

**Table 11:** Comparison of the effective area measured for the flight spare telescope at three energies compared with the analytic combination of the independent MMA and detector calibrations.

| Energy (keV) | Effective Area-1 ($cm^2$) | Effective Area= 2 ($cm^2$) |
|---|---|---|
| **2.3** | 25.38±5% | 26.02±3% |
| **4.6** | 7.37±4% | 7.56±3% |
| **6.6** | 2.81±2% | 2.68±3% |

### 7.3.2. Angular Resolution

For the telescope, the HPD was measured at the same energies as was the MMA, using the same 8-mm-diameter circular region on the DU. As with the MMA, the diameter containing half the flux within this region was calculated and converted to an angle.

It is expected that the telescope and MMA HPDs will differ slightly because the DU adds additional blurring to the image due to its finite spatial resolution (limited by the ability to determine the beginning of a track) and by the finite depth of the detector (10 mm), which adds additional defocusing of the image due to gas transparency and the cone angle of X rays focused by the mirror assembly. However, these two components are small compared to the native MMA resolution.

Defocusing effects, due to finite detector gas depth, were calculated via Monte-Carlo simulations. As a check, results were also obtained by taking a series of CCD images over a ± 5-mm region centered on the optimum focal distance, to span the region covered by the thickness of the detector gas cell. Summing these CCD images effectively blurs the image in a similar way to the gas depth in the DU. These measurements agreed very well with the Monte-Carlo simulations (within a fraction of an arcsecond).

Table 12 shows the results of the HPD comparison. MMA HPD represents the angular resolution of the MMA measured with the CCD camera. Detector spatial resolution and defocusing effects, converted to HPD in arcsec, are summed in quadrature with the native MMA resolution to give the predicted telescope angular resolution. The final column shows the measured telescope HPD obtained during telescope calibration. Estimated uncertainties are ~ 2% (1σ) for the predicted HPD measurements and ~ 0.5-1% for the measured value. As expected, good agreement between predicted and measured telescope response is obtained.

**Table 12**: Comparison of predicted and measured on-axis telescope angular resolution

| X-ray Tube (Line) | MMA HPD (arcsec) | Detector Spatial Resolution (arcsec) | Detector Defocusing Effects (arcsec) | Predicted Telescope HPD (arcsec) | Measured Telescope HPD (arcsec) |
|---|---|---|---|---|---|
| 2.3 keV | 20.0 | 5.6 | 6.0 | 21.6 | 22.2 |
| 4.6 keV | 20.8 | 6.3 | 7.0 | 22.8 | 23.8 |
| 6.6 keV | 20.1 | 7.4 | 7.2 | 22.6 | 24.1 |

### 7.3.3. Modulation Factor

Telescope modulation factors were measured for three source + crystal combinations (see Table 3), with source energies and crystal lattice spacings chosen to achieve Bragg angles close to 45°. Consequently, the resulting x-ray beams are nearly totally (linearly) polarized.

For all measurements of modulation factors, dithering of the MMA was used to simulate the dithering that will be used on orbit. This dithering also approximately aligns with that used for detector calibrations in Italy. The actual dithering pattern was a Lissajous-type figure, designed to provide near uniform coverage of the dither area, with a radius of 3.6 mm.

Spurious modulation, measured for each energy during calibration of this detector, was first removed. Then the DU data is processed using a standard moments analysis with a 20% cut on tracks, based on track ellipticity, and an energy cut of ± 3-σ around all line energies. The measured modulation factors for the telescope are then compared to those measured for the DU alone during the spare detector calibration performed by IAPS in Italy. The results are shown in Table 19, together with the 1-σ measurement errors.

**Table 13:** Measured telescope and DU modulation factors

| Source / Energy | Telescope Modulation Factor Spare DU + Spare MMA | Detector Modulation Factor Spare DU @ IAPS |
|---|---|---|
| **Rh L (2.7 keV)** | 29.77 +/- 0.13 % | 29.87 +/- 0.13 % |
| **Ti K (4.5 keV)** | 46.18 +/- 0.21 % | 46.04 +/- 0.14 % |
| **Fe K (6.4 keV)** | 56.26 +/- 0.23 % | 56.59 +/- 0.09 % |

Table 13 shows that modulation factors measured with the telescope are statistically identical (within 1-σ) to those measured with the DU alone. This confirms that for the modulation factor the telescope response can be derived from the DU calibration data alone, and that there are no MMA-induced effects.

## 7.4. Alignment

Each IXPE focal plane detector has an active area of 15 mm ×15 mm and so each telescope must be carefully aligned to ensure the image is near the center of the detector and that the axis of each mirror module is co-aligned with the star tracker. The driving requirement here is the capability to simultaneously place an extended source of diameter up to 9 arcminutes on all three detectors.

Alignment primarily consisted of placing the three DUs and the three MMAs on congruent triangles, the former on the top deck of the spacecraft and the latter on the deployable mirror module support structure. Surface mount reflectors (SMRs) installed on the MMAs and DUs during construction enabled this process accomplished using a laser tracker system. Precise knowledge linking the SMR positions to the respective nodes of the MMAs and the DUs was derived to high precision (< 100 µm) during component assembly. Before the MMAs were finally placed in position, their X-ray axes were co-aligned with each other and with the optical axis of the forward star tracker using alignment cubes.

The deployment accuracy of the boom is such that the X-ray image of an on-axis X-ray source should be within ~ 1 mm (1-σ) of the center of each detector in the X, Y, and Z axes. For precise alignment, use may be made of the Tip/Tilt/Rotate (TTR) system to make adjustments on-orbit in the X and Y axes. Situated between the boom and the mirror module support structure, the TTR can effectively repoint the observatory to move the image on the detector which then moves to re-acquire the target (see Figure 16).

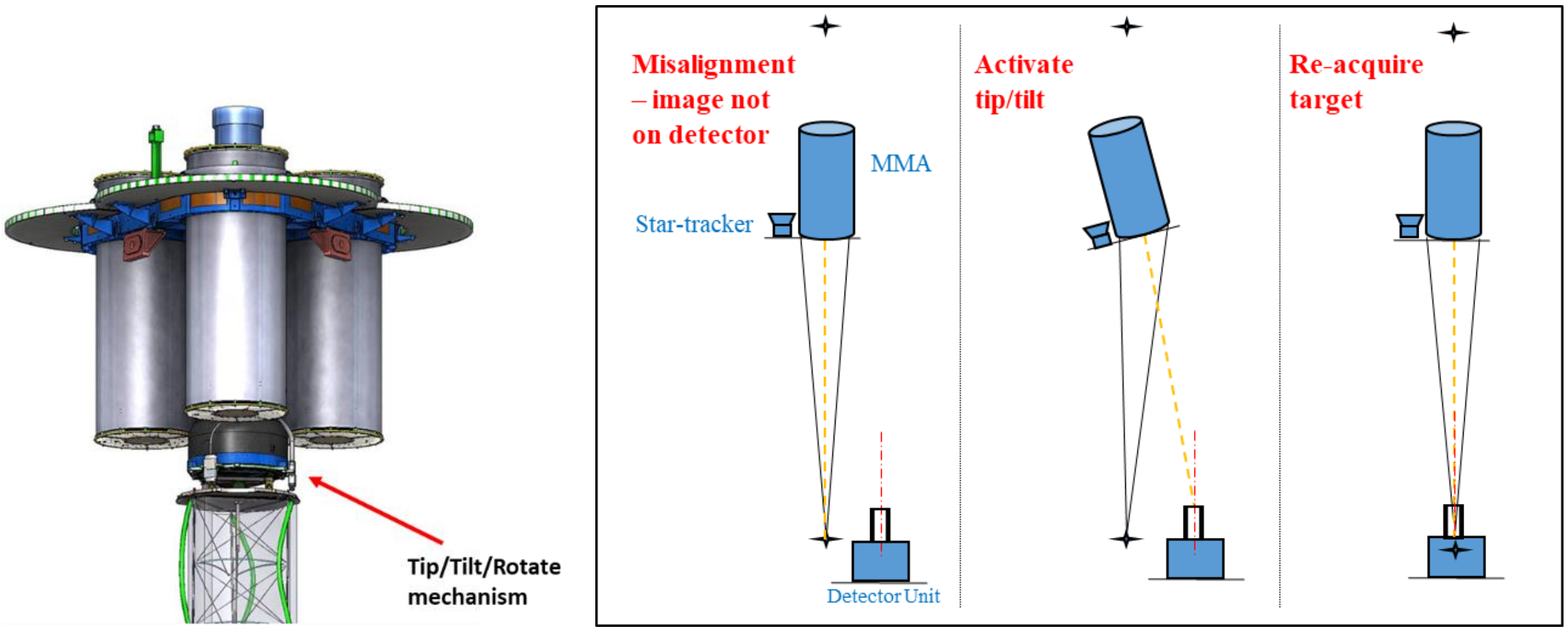

**Figure 16**: Left image shows the Tip/Tilt/Rotate mechanism; Right image shows the on-orbit alignment process

## 7.5. In-Flight Calibration

To enable in-flight calibration monitoring, each DU is equipped with a filter and calibration wheel assembly [19] (Figure 17). These assemblies contain various radioactive sources that can be rotated in front of the GPD to provide for monitoring gain, energy resolution, spurious modulation, and the modulation factor. The calibration sources will be used in those parts of the orbit when the X-ray source under study is eclipsed by the earth, calibrating one detector at a time.

The calibration sources are all based on $^{55}Fe$ isotopes which have a $K_\alpha$ line at 5.9 keV and a $K_\beta$ line at 6.5 keV. Cal source A produces polarized X rays at 3 keV (via a silver target, Si $L_\alpha$ = 3 keV) and at 5.9 keV, through 45° Bragg reflection off a graphite mosaic crystal. Cal sources B and C have unpolarized 5.9 keV and 6.5 keV X rays in a spot (~3mm) and a "flood" (~ 15 ×15 mm) configuration. Cal source D utilizes a silicon target in front of the $^{55}Fe$ to produce a broad beam at 1.7 keV (Si $K_\alpha$).

In addition to the calibration sources, the wheel also contains an open position, a closed position and an attenuator position consisting of a 75-µm-thick Kapton foil coated with 100 nm of aluminum on each side. The first of these is for normal operations, the second for internal background measurements and the third is for observing very bright sources which would otherwise exceed the throughput of the system

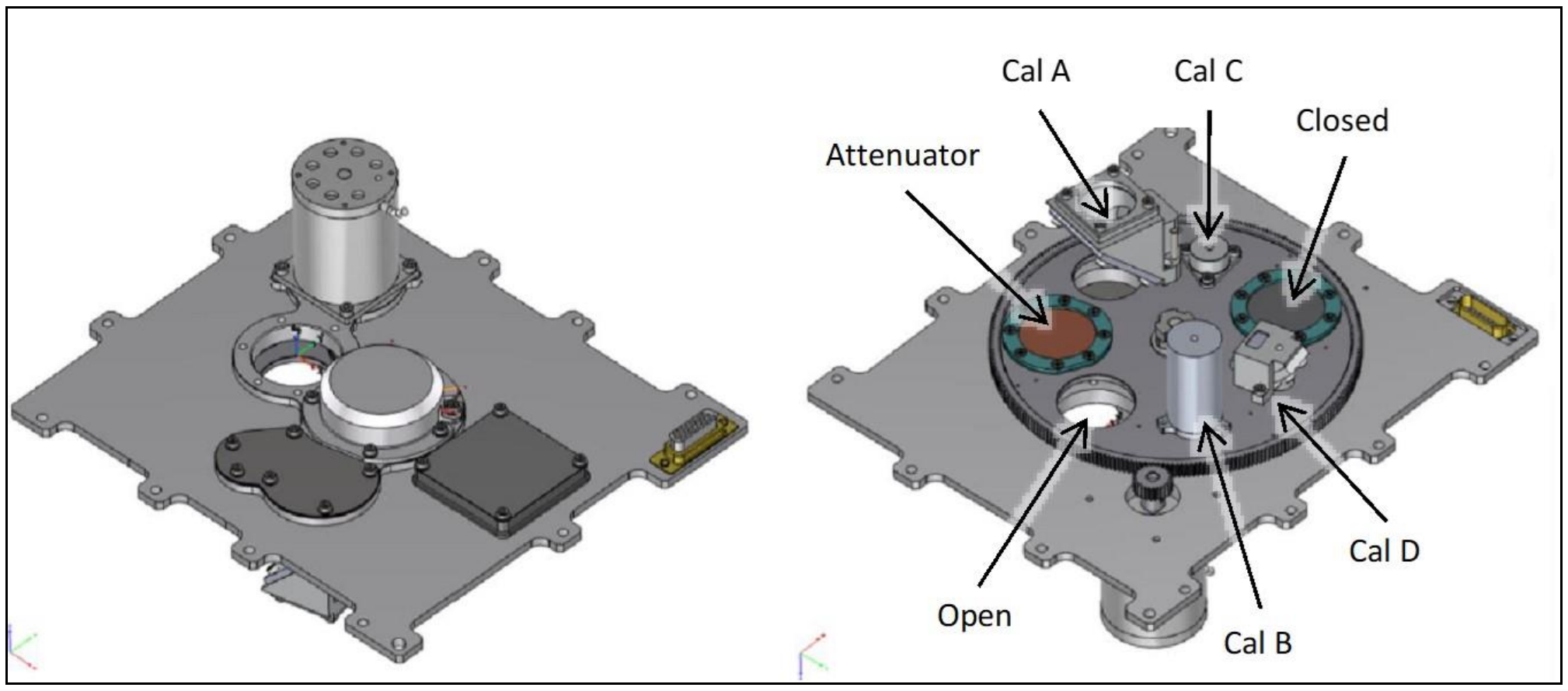

**Figure 17:** Filter and calibration wheel assembly (left: top view; right: bottom view)

## 8. IXPE SCIENCE

While several bright polarized point sources (e.g. accretion-powered binaries) should provide high quality polarization data, the imaging “I” in IXPE puts a special emphasis on spatially resolvable sources. Of course, the classical soft X-ray standard is the Crab pulsar and its nebula and IXPE’s 30″ HPD allows a coarse map of the nebula torus and jets along with a phase-resolved study of the central pulsar. Averaging these polarization images permits a reasonable comparison with the classic OSO-8 measurement [8] after allowing for the expansion of the nebula as well as the decrease in available energy from the pulsar. A few other PSR and pulsar wind Nebulae (PWNe) are also accessible, with PSR B1509-85/MSH 15-52 being a particularly attractive target as the power-law (i.e. polarized synchrotron) soft X-ray emission extends across the 13′ IXPE field of view.

However, the most striking early images are likely to come from the shell-type supernova remnants, Cas A (Figure 18), Tycho and SN 1006. The first two are well-matched to the IXPE field-of-view while for SN1006, we will focus on bright limb regions. Again power-law X-rays indicate polarized synchrotron emission which even dominates in some regions of the limb forward shock – but the bulk of detected photons will be thermal. Here IXPE’s modest energy and angular resolution are essential to extracting the regions and energy bands with the brightest synchrotron signal. While the detailed polarization level is unknown, predictions based on radio polarization maps suggest that we should detect polarization in a dozen independent regions and, coherently averaging, measure large scale polarization patterns to even fainter flux levels. Since the polarization direction and level probe the acceleration zone magnetic field geometry and coherence, we can study field amplification and feedback in the particle acceleration zones.

Other Galactic sources are of great interest, as well. One project uses the scattering induced accretion disk polarization and its propagation through curved space to investigate the spin of the black holes in bright binary systems such as GRS1915+105 and Cygnus X-1. Here the energy dependence maps to disk radial position and the Kerr-metric induced spread of the polarization vector allows a novel measurement of curved space to be compared with other BH X-ray spin constraints. The accreting neutron stars are, of course, expected to provide a rich harvest of variable polarization signals. Her X-1, for example, allows the study of the magnetic accretion column at a variety of spin-, orbit- and precession-phases. For targets such as the 1.7 ms accreting pulsar IGR J00291+5934, IXPE’s excellent time tagging will be essential to the phase-resolved studies. Note that this object and many other accretion-powered neutron star and black hole targets are transient; IXPE will use external (e.g. SWIFT) flux monitors to trigger their observations. It is anticipated that such targets of opportunity will be added to the observing plan and executed on a 48 hr timescale. In the bright state such targets often rapidly saturate on-board event storage and so short observations interleaved with lower-flux targets will allow downloads to keep up with the data stream. Interestingly, much of the polarization modeling for bright accretion-powered pulsars stems from the 70’s and 80’s; with the rich expected IXPE polarization phenomenology we already see a renaissance in such modeling efforts.

Another class of Galactic sources, where polarization modeling is key, are the magnetars where the neutron stars $10^{14}$-$10^{15}$G magnetic fields induce novel propagation effects involving quantum electrodynamics (QED). Early planned observations include RXS J170849.0-400910 and 4U 0412+61. Long exposures are needed to build up high statistics polarization studies of their pulsed emission, and the presence of vacuum birefringence is modeled to strongly affect the pulse polarization [20] (while hardly affecting the total intensity). In this way, IXPE observations of these sources can be used to demonstrate the effect of this important QED phenomenon.

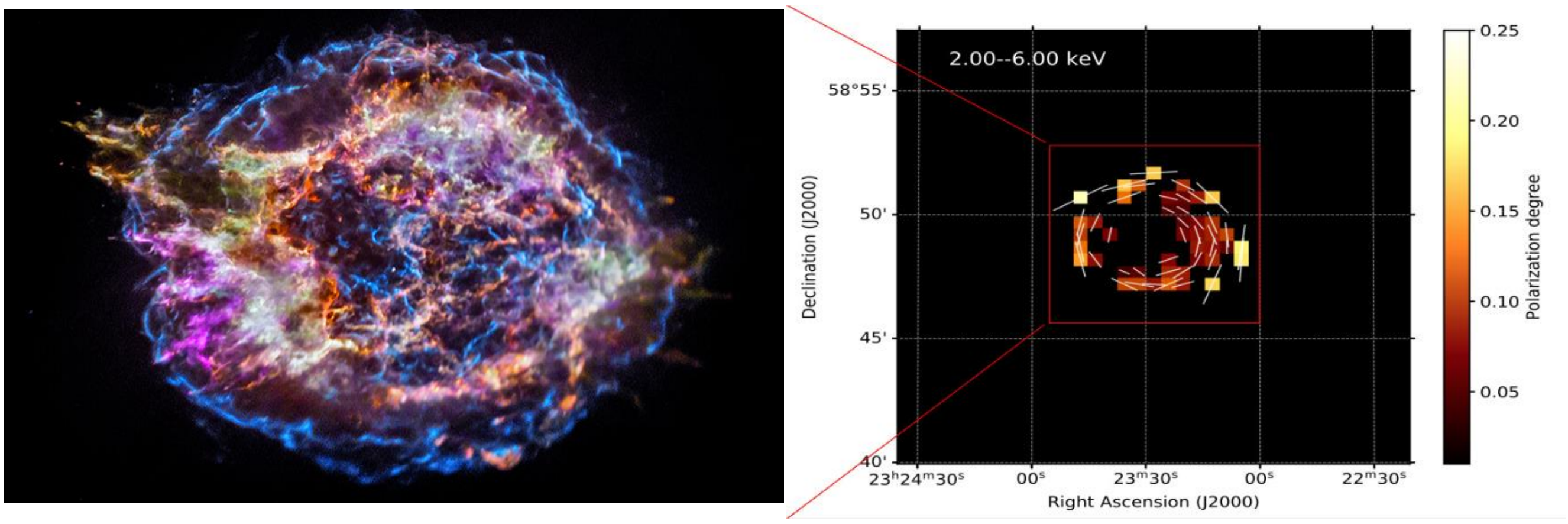

**Figure 18:** Left: a Chandra X-ray Observatory CCD image of Cas A. Right: IXPE pixels delivering >3σ 2-6keV polarization measurements assuming 20% tangential polarization of the non-thermal flux and a 1Ms exposure

Turning to the massive black hole domain, IXPE is expected observe several active galaxies, aiming to measuring the geometry of their X-ray emission regions and the circumnuclear matter. One target of particular interest is the (presently very weakly active) Galactic center. Here scattered light signals from molecular clouds near Sgr A* indicates much higher X-ray luminosity in the past. IXPE measurements of the polarized flux and position angle can lock down the scattering geometry, testing whether past outbursts of Sgr A* are indeed the origin of the high luminosity events and even time-delay date the outbursts (Figure 19). The Galactic center X-ray clouds are extended and of low surface brightness, so IXPE's imaging capabilities and the relatively low particle background enabled by its near-equatorial orbit are definitely needed for a successful measurement. Even so, these observations are challenging, requiring exposures of 1-2 Ms.

Several relativistic jet sources, principally the so-called `blazars', will also be examined with IXPE. Here, the X-ray band samples either the upper end of the electron synchrotron emission component (for the High Synchrotron Peakss like Mrk 421 and Mrk 501) or the high energy peak component, likely electron Compton emission, but plausibly proton synchrotron emission (for the Low Synchrotron Peaks like 3C 454.3 and 3C 279). Polarization is a strong diagnostic of the emission processes in both cases and comparison with lower energy (mm, optical) polarization measurements and with multi-wavelength (radio-TeV) flaring activity tests jet acceleration and radiation models. In fact, a few targets appear to spectrally transition between the two cases right in the IXPE energy range and thus study of their polarization versus energy can probe the extremum of the jet acceleration zone.

IXPE polarization measurements thus offer new physics probes of a variety of objects. Current plans suggest that over a two-year prime mission we can schedule observations of ~100 steady targets and 10-20 transients, offering a first polarized look at many classes of X-ray sources. We anticipate that these discovery observations will only whet astrophysicist's appetites, resolving ambiguities left from basic intensity measurements and driving significant advances in the theoretical modeling of high energy sources. An extended mission with a guest observer program could offer the opportunity to push the envelope even further, with very deep exposures measuring new source types of particular interest and coordinated multi-wavelength (or even multi-messenger) campaigns giving new physics insights. But IXPE is very much an "explorer" and we anticipate that a successful mission will only strengthen the drive for a future polarization facility with improved energy range, angular resolution and, especially, the large effective area needed to fully exploit X-ray polarization as a new astrophysics tool.

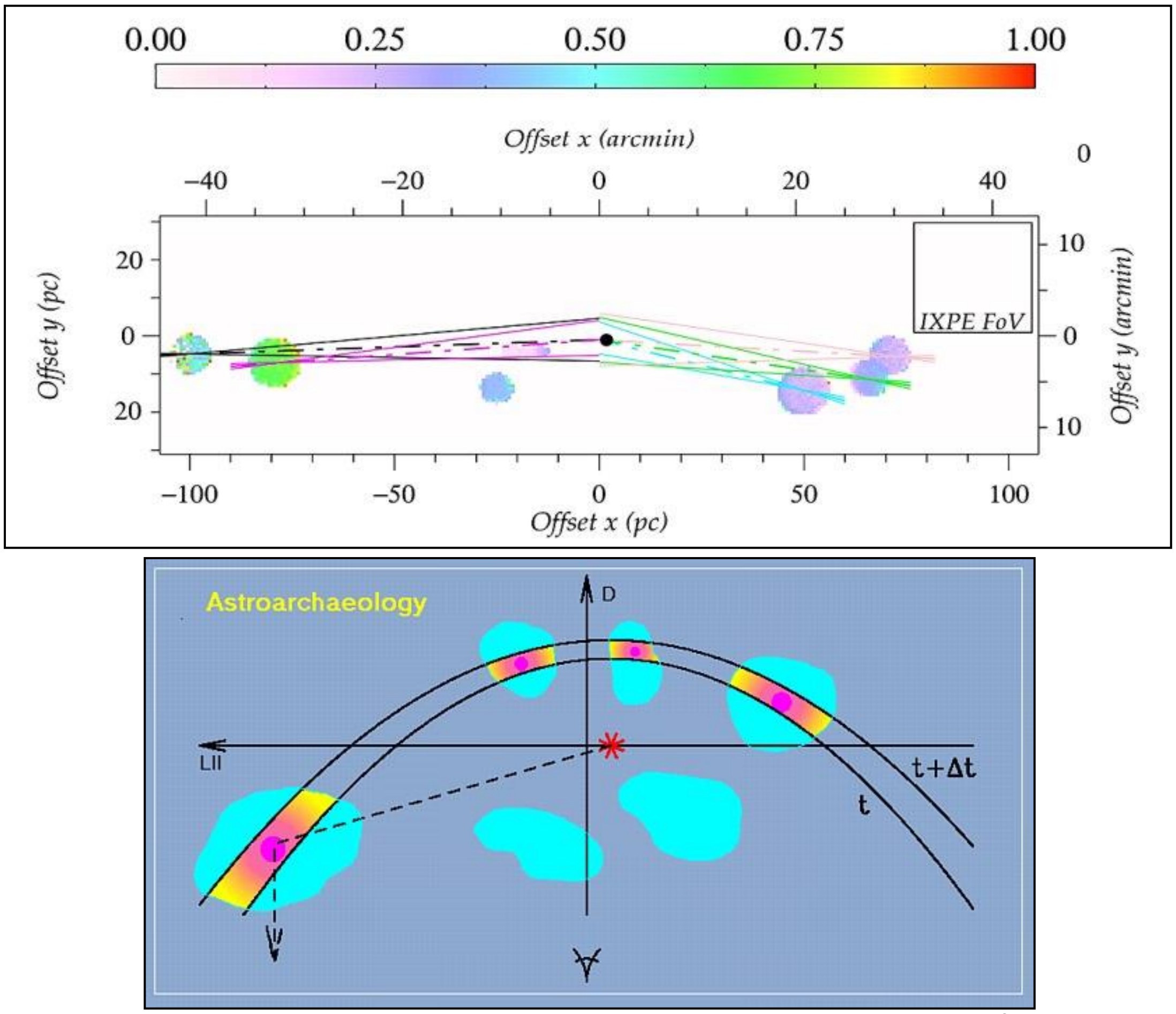

**Figure 19:** Galactic center schematic (upper), showing a time-delayed scattering signal from the clouds (lower).

[12] Muleri, F., Piazzolla, R., Di Marco, A., Fabiani, S., La Monaca, F., Lefevre, C., Morbidini, A., Rankin, J., Soffitta, P., Tobia, A., Xie, F., Amici, F., Attinà, P., Bachetti, M., Brienza, D., Centrone, M., Costa, E., Del Monte, E., Di Cosimo, S., Di Persio, G., Evangelista, Y., Ferrazzoli, R., Loffredo, P., Perri, M., Pilia, M., Puccetti, S., Ratheesh, A., Rubini, A., Santoli, F., Scalise, E., Trois A., "The IXPE instrument calibration equipment", Astroparticle Physics, Volume 136, March 2022, 102658

[13] Rankin, J., Muleri, F., Tennant, A., Bachetti, M., Costa, E., Di Marco, A., Fabiani, S., La Monaca, F., Soffitta, P., Tobia, A., Trois, A., Xie, F., Baldini, L., Di Lalla, N., Manfreda, A., O'Dell, S., Perri, M., Puccetti, S., Ramsey, B., Sgro', C., Weisskopf, M., "An algorithm to calibrate and correct the response to unpolarized radiation of the X-ray polarimeter on board IXPE", 2021 Accepted for publication in the Astronomical Journal.

[14] Fabiani, S., Di Marco., A., La Monaca, F., Muleri, F., Rankin, J., Soffitta, P., Xie, F., Amici, F., Attinà, P., Bachetti, M., Baldini, L., Barbanera, M., Baumgartner, W., Bellazzini, R., Borotto, F., Brez, A., Brienza, D., Caporale, C., Cardelli, C., Carpentiero, R., Castellano, S. Castronuovo, M., Cavalli, L., Cavazzuti, E., Ceccanti, M., Centrone, M., Citraro, S., Costa, E., D'Alba, E., D'Amico, F., Del Monte, E., Di Cosimo, S., Di Lalla, N., Di Persio, G., Donnarumma, I., Evangelista, Y., Ferrazzoli, R., Latronico, L., Lefevre, L., Loffredo, L., Lorenzi, P., Lucchesi, L., Magazzù, C., Magazzù, G., Maldera, S., Manfreda, A., Mangraviti, E., Marengo, M., Giorgio Matt, Mereu, P., Minuti, M., Morbidini, A., Mosti, F., Nasimi, H., Negri, B., Nuti, A., O'Dell, S., Orsini, L., Perri, M., Pesce-Rollins, M., Piazzolla, R., Pieraccini, S., Pilia, M., Pinchera, M., Profeti, A., Puccetti, S., Ramsey, B., Ratheesh, A., Rubini, A., Santoli, F., Sarra, P., Scalise, E., Sciortino, A., Sgrò, C., Spandre, G., Tardiola, M., Tennant, A., Tobia, A., Trois, A., Vimercati, M., Weisskopf, M., Zanetti, D., Zanetti, F. "Measurement of the quantum efficiency of the IXPE Detector Units" in preparation (2021) .

[15] La Monaca, F., Fabiani, S., Lefevre, C., Morbidini, A., Piazzolla, R., Amici, F., Attinà, P., Brienza, D., Costa, E., Di Cosimo, S., Di Marco, A., Di Persio, G., Evangelista, Y., Ferrazzoli, R., Loffredo, P., Fabio Muleri, John Rankin, Ajay Ratheesh, Alda Rubini, Francesco Santoli, Emanuele Scalise, Paolo Soffitta, Tobia, A., Trois, A., Xie, F., Zeiger, B., "IXPE DU-FM ions-UV filters characterization", Proc. SPIE 11444, Space Telescopes and Instrumentation 2020: Ultraviolet to Gamma Ray, 1144463 (12 January 2021).

[16] Di Marco, A., Fabiani, S., La Monaca, F., Muleri, F., Rankin, J., Soffitta, P., Xie, F., Amici, F., Attinà, P., Bachetti, M., Baldini, L., Barbanera, M., Baumgartner, W., Bellazzini, R., Borotto, F., Brez, A., Brienza, D., Caporale, C., Cardelli, C., Carpentiero, R., Castellano, S. Castronuovo, M., Cavalli, L., Cavazzuti, E., Ceccanti, M., Centrone, M., Citraro, S., Costa, E., D'Alba, E., D'Amico, F., Del Monte, E., Di Cosimo, S., Di Lalla, N., Di Persio, G., Donnarumma, I., Evangelista, Y., Ferrazzoli, R., Latronico, L., Lefevre, L., Loffredo, L., Lorenzi, P., Lucchesi, L., Magazzù, C., Magazzù, G., Maldera, S., Manfreda, A., Mangraviti, E., Marengo, M., Giorgio Matt, Mereu, P., Minuti, M., Morbidini, A., Mosti, F., Nasimi, H., Negri, B., Nuti, A., O'Dell, S., Orsini, L., Perri, M., Pesce-Rollins, M., Piazzolla, R., Pieraccini, S., Pilia, M., Pinchera, M., Profeti, A., Puccetti, S., Ramsey, B., Ratheesh, A., Rubini, A., Santoli, F., Sarra, P., Scalise, E., Sciortino, A., Sgrò, C., Spandre, G., Tardiola, M., Tennant, A., Tobia, A., Trois, A., Vimercati, M., Weisskopf, M., Zanetti, D., Zanetti, F., "Calibration of modulation factor of IXPE focal plane X-ray polarimeters" in preparation (2021).

[17] Di Marco, A., Costa E., Muleri, F., Soffitta, P., Fabiani, S., La Monaca, F., Rankin, J., Xie, F., Bachetti, M., Baldini, L., Baumgartner, W., Bellazzini, R., Brez, A., Castellano, S., Del Monte, E., Di Lalla, N., Ferrazzoli, R., Latronico, L., Maldera, S., Manfreda, A., O'Dell, S., Perri, M., Pesce-Rollins, M., Puccetti, S., Ramsey, B., Ratheesh, A., Sgro', C., Spandre, G., Tennant, A., Tobia, A., Trois, A., Weisskopf, M. "A weighted analysis to improve the X-ray polarization sensitivity of IXPE" submitted to the Astronomical Journal (2021).

[18] Xie, F., Ferrazzoli, R., Soffitta, P., Fabiani, S., Costa, E., Muleri, F., Di Marco, A., "A study of background for IXPE", Astroparticle Physics 128 (2021).

[19] Ferrazzoli, R., Muleri, F., Lefevre, C., Morbidini, A., Amici, F., Brienza, D., Costa, E., Del Monte ,E., Di Marco, A., Di Persio, G., Donnarumma, I., Fabiani, S., La Monaca, F., Loffredo, P., Maiolo, L., Maita, F., Piazzolla, R., Ramsey, B., Rankin, J., Ratheesh, A., Rubini, A., Sarra, P., Soffitta, P., Tobia, A. & Xie, F., "In-flight calibration system of imaging x-ray polarimetry explorer", J. Astron. Telesc. Instrum. Syst., Vol 6(4), (2020).

[20] Taverna, R., Turolla, R., Suleimanov, V., Potekhin, A. Y., Zane, S. "X-ray spectra and polarization from magnetar candidates", MNRAS, Volume 492, Issue 4, p.5057-5074 (March 2020).